\documentclass[useAMS,usegraphicx]{mn2e}
\input{psfig}
\usepackage{times}
\usepackage[usenames]{color}
\definecolor{Blue}{rgb}{0,0.08,0.65}
\definecolor{Red}{rgb}{0.65,0.08,0.05}
\definecolor{Green}{rgb}{0.15,0.45,0.25}

\newif\ifAMStwofonts

\def\build#1_#2^#3{\mathrel{
\mathop{\kern 0pt#1}\limits_{#2}^{#3}}}
\def\ga{\mathrel{\mathchoice {\vcenter{\offinterlineskip\halign{\hfil
$\displaystyle##$\hfil\cr>\cr\sim\cr}}}
{\vcenter{\offinterlineskip\halign{\hfil$\textstyle##$\hfil\cr>\cr\sim\cr}}}
{\vcenter{\offinterlineskip\halign{\hfil$\scriptstyle##$\hfil\cr>\cr\sim\cr}}}
{\vcenter{\offinterlineskip\halign{\hfil$\scriptscriptstyle##$\hfil
\cr>\cr\sim\cr}}}}}
\def\la{\mathrel{\mathchoice {\vcenter{\offinterlineskip\halign{\hfil
$\displaystyle##$\hfil\cr<\cr\sim\cr}}}
{\vcenter{\offinterlineskip\halign{\hfil$\textstyle##$\hfil\cr<\cr\sim\cr}}}
{\vcenter{\offinterlineskip\halign{\hfil$\scriptstyle##$\hfil\cr<\cr\sim\cr}}}
{\vcenter{\offinterlineskip\halign{\hfil$\scriptscriptstyle##$\hfil
\cr<\cr\sim\cr}}}}}
\begin{document}
\title[Accurate estimators of power spectra in $N$-body simulations]{Accurate estimators of power spectra in $N$-body simulations}
\author[Colombi, Jaffe, Novikov \& Pichon]{
  St\'ephane
  Colombi,$^1$\thanks{E-mails: colombi@iap.fr (SC),
    a.jaffe@imperial.ac.uk (AJ), d.novikov@imperial.ac.uk (DN),
    pichon@iap.fr (CP).  } 
  Andrew Jaffe,$^{2\star}$
  Dmitri Novikov,$^{2\star}$
  Christophe Pichon,$^{1\star}$\\
\\
   $^1$ Institut d'Astrophysique de Paris, UMR7095 CNRS, Univ. P. \& M. Curie, 98 bis Boulevard Arago, 
			75014 Paris, France\\
   $^2$ Astrophysics, Blackett Laboratory, Imperial College London, London SW7 2AZ}
\maketitle
\begin{abstract}
A method to rapidly estimate the Fourier power spectrum of a
point distribution is presented. This method relies on a Taylor
expansion of the trigonometric functions. It yields the Fourier modes from a number of FFTs, which is controlled by the order $N$ of the expansion 
and by the  dimension $D$ of the system.
In three dimensions, for the practical value $N=3$, the number of FFTs required is 20.

We apply the method to the measurement of the power spectrum
of a periodic point distribution that is a local Poisson realization
of an underlying stationary field. We derive explicit analytic
expression for the spectrum, which allows us to quantify---and correct for---the biases induced by discreteness and by the truncation of the Taylor expansion, and to bound
the unknown effects of aliasing of the power spectrum.
We show that these aliasing effects decrease rapidly with the  order $N$.
For $N=3$, they are expected to be respectively smaller than 
$\sim 10^{-4}$ and $0.02$ at half the Nyquist frequency and at 
the Nyquist frequency of the grid used to perform the FFTs.
The only remaining significant source of errors is
reduced to the unavoidable cosmic/sample variance due to the finite 
size of the sample.

The analytical calculations are successfully
checked against a cosmological $N$-body experiment. We
also consider the initial conditions of this simulation,
which correspond to a perturbed grid. This allows us to test a case
where the local Poisson assumption is incorrect. Even in that
extreme situation, the third-order Fourier-Taylor estimator behaves
well, with aliasing effects restrained to at most the percent
level at half the Nyquist frequency.

We also show how to reach arbitrarily large dynamic
range in Fourier space (\emph{i.e.}, high wavenumber), while keeping statistical
errors in control, by appropriately ``folding'' the particle
distribution.
\end{abstract}
\begin{keywords}
methods: analytical, data analysis, numerical, statistical, $N$-body
simulations -- cosmology: large-scale structure of Universe
\end{keywords}
\section{Introduction}
The power spectrum, $P(k)$, represents the primary tool to characterize
the clustering properties of the large scale structure of the
universe. Most of major constraints on cosmological models and on
cosmological parameters have been derived from measuring $P(k)$
or its Fourier transform, the two-point
correlation function. For instance, the tight constrains derived from
WMAP experiment rely on measurements of the power spectrum in
spherical harmonic space (\emph{e.g.}, Dunkley et al., 2008);
the most significant results from weak lensing
analysis come from measurements of the two-point correlation function of the
cosmic shear (\emph{e.g.}, Benjamin et al., 2007; Fu et al., 2008); 
the analysis of the power spectrum of
absorption lines of lyman-$\alpha$ forest allowed one to
infer drastic constraints on the clustering properties of the matter
distribution at small scales (\emph{e.g.}, Croft et al., 1999); 
and, last but not least, the two-point correlation
function and the power spectrum have been used extensively to analyse
directly the clustering properties of 2 and 3 dimensional galaxy catalogs 
(\emph{e.g.}, Peebles, 1980; Baumgart \& Fry, 1991; Martinez, 2008, 
for a recent general review on the subject). 

To be able to derive predictions from models of large scale
structure formation, there has been successful attempts to find
universal dynamical laws, partly phenomenological, that lead to
semi-analytical expressions of the non linear power spectrum
(or the two-point correlation function) of the matter distribution. 
Among them, one can cite 
the nonlinear ansatz of Hamilton et al.~(1991), later improved by Peacock
\& Dodds (1996, see also Smith et al., 2003). Such a non-linear ansatz has been
used to constrain  models against observations,
particularly in weak lensing surveys (\emph{e.g.}, Benjamin et
al. 2007; Fu et al. 2008). 
Another well known phenomenological description is the so called
halo model, which proposes not only  some insights on the clustering
properties of the dark matter distribution, but also of
the galaxy distribution itself  (see, \emph{e.g.}, Ma \& Fry, 2000; 
Peacock \& Smith, 2000; Seljak, 2000; Scoccimarro et al. 2001; 
see Cooray \& Sheth 2002 for an extensive review). 
Again, the measurement of two-point statistics in the galaxy
distribution was used to infer important constraints
on the halo model parameters  (\emph{e.g.}, Abazajian et al. 2005).

With the advent of very large modern surveys,
``precision cosmology'' has become a reality: the accuracy of the
observations have caught up to the accuracy of the predictions. These
need to be more and more precise to constrain
current models of large scale structure formation, for instance
the fiducial $\Lambda$CDM (Cold Dark Matter) model. It is
therefore now crucial to obtain very fine estimates of statistics
in simulations with good control of the systematic errors in
order to be able to fine tune non linear ansatz such that of Hamilton et al. 
(1991) or details on the set up of the halo model.

In this work, we concentrate on the problem of measuring as accurately
as possible the Fourier modes of a distribution of points such as those
coming from an $N$ body simulation, with a particular emphasis
on the power spectrum.  The traditional method for measuring 
the Fourier modes, $\delta_k$, consists in
assigning the point distribution to a periodic grid with some interpolation
method and then computing $\delta_k$ with a Fast Fourier
Transform (FFT) technique. However, the introduction  of a grid,
combined with the corresponding interpolation, induces two effects: damping of the modes
at large $k$ due to the convolution involved in the interpolation, and
effects of aliasing due  to the finite resolution of the
grid (\emph{e.g.}, Hockney \& Eastwood, 1988). 
In addition, the discrete nature of the particle distribution induces
some systematic effects, but these latter can be straightforwardly 
accounted for if the distribution of particles is a local Poisson
process (\emph{e.g.}, Peebles, 1980).
While the bias induced by the interpolation method can also be easily corrected  
for, the effects of aliasing are more difficult to control.
This has been for instance illustrated well by Jing (2005),
who proposed to correct for aliasing using an iterative method, based 
on an ansatz that assumes that the power spectrum behaves like a
power law at large $k$. However this method, although efficient
for cosmological power spectra which behave close to power laws,
is not free of biases in general. 
An alternative route involves using appropriate interpolation functions,
which are by construction meant  to reduce  the
effects of aliasing as much as possible. 
This is for example the case of the 
Daubechies wavelets (Daubechies, 1988), as proposed
by Cui et al.~(2008). While these interpolating functions are
powerful, there are still some significant residuals at the few
percent level, and Cui et al.~(2008) do not provide a rigorous
way to quantify  and correct them. 

The aim of this paper is more ambitious than Jing (2005) and Cui et
al. (2008): we want to be able to measure the power spectrum from a
simulation at  an arbitrary level of accuracy, with rigorous control
of the biases and the residuals due to aliasing. Of course, the higher
the required level of accuracy, the larger the computational
cost. Furthermore, even though we shall be able to measure the power spectrum  extremely
accurately from a given simulation, it does not
mean that the power spectrum of the underlying cosmology will be
estimated fairly: a statistical error --- cosmic variance ---
arising  from the finite number of available modes given the 
finite size of the simulated volume will still be present  
(\emph{e.g.}, Feldman, Kaiser \& Peacock, 1994;
Scoccimarro, Zaldarriaga \& Hui, 1999; Szapudi, 2001; Bernardeau et al. 2002 for
a review).

The method we propose is inspired by Anderson \& Dahleh
(1996). It is based on the fact that the
Taylor series expansion of trigonomic functions, $\sin(x)$ and $\cos(x)$,
converges very rapidly. This will allow us to compute
Fourier modes efficiently with a number of FFT depending
on the order $N$ of the Taylor series expansion used. That number
will control the effects of aliasing as well as the biases on
the rough estimator, which can be corrected for in the case of a local
Poisson realization of a stationary random field. We shall
write explicit analytical expressions for the power spectrum
and propose an unbiased estimator that will be tested 
in detail against a controlled $N$-body experiment.

This paper is organized as follows. First we describe what we call the
Fourier-Taylor transform and its practical implementation
(\S~\ref{sec:Fouriertaylor}). Then, we
construct a rough estimator of the power spectrum from it, and perform
analytical calculation of its ensemble average by 
assuming local Poisson sampling of a stationary
random field (\S~\ref{sec:ftps}). 
This section is supplemented with
Appendix A, which discusses some subtle differences between the
unconstrained versus the constrained ensemble average, and Appendix
B, which details some useful analytic expressions of various
quantities occurring in the calculations.
We study the biases on the rough estimator of the 
power spectrum, which
can be easily corrected for,  as well as
the unknown residuals due to aliasing, 
which are controlled by the order of the Taylor expansion. 
The analytic results are then validated in a CDM
{\tt GADGET} $N$-body simulation (\S~\ref{sec:validation}).
We study two configurations, the final stage of the simulation,
which should agree very well with the assumption of local Poisson
sampling and  stationarity, and the initial conditions, corresponding
to a slightly perturbed grid. This section is supplemented with
Appendix C, which details  the calculation of the power spectrum of a randomly
perturbed grid.
In \S~\ref{sec:extension}, we show how to cover all the
available dynamic range in Fourier space while keeping
control of the errors, by appropriate foldings of
the particle distribution.  
Finally, section~\ref{sec:conclusion}
briefly summarizes the results of this paper.
\section{The Fourier-Taylor transform}
\label{sec:Fouriertaylor}
We begin with a discrete distribution of points $x_i$ with weights
(masses) $M_i$, $i=1,\ldots, N_{\rm p}$, where each $x_i$ is
potentially a $D$-dimensional vector. The equivalent perturbed density field is 
\begin{equation}
	\rho(x) \equiv \frac{1}{N_{\rm p}}\sum_{i=1}^{N_{\rm p}} M_i\ \delta_D(x-x_i)
\end{equation}
where $\delta_D(x)$ is the $D$-dimensional Dirac delta function.
The Fourier transform of this distribution is 
\begin{eqnarray}
\delta_k &\equiv& \int d^Dx\; \rho(x)\ e^{I\; k\cdot x}\nonumber\\
&=&\frac{1}{N_{\rm p}} \sum_{i} M_i \exp(I k\cdot x_i),
\label{eq:distra}
\end{eqnarray}
where $k$ is a $D$-dimensional wavevector and the imaginary unit is $I^2=-1$ (we use lower case $i$ as an integer index).
Since the number of dimensions is assumed to be
arbitrary the operator ``$\cdot$" is the
scalar product. The direct calculation of the sum in equation~$(\ref{eq:distra})$
is a very slow, an $N_k \times N_{\rm p}$ process, where $N_k$ is the number of
sought wavenumbers. 

To speed up the calculation, one can choose a homogeneous cubic grid of a
certain size covering the volume occupied by all the points, 
$N_{\rm g}$ (in 3D, $N_{\rm g} \times N_{\rm g} \times
N_{\rm g}$),\footnote{The following calculations can be easily
generalized to a rectangular grid.} and define the function $J(i)$ giving
the (vector) integer position of the cell containing object $i$. Then equation~(\ref{eq:distra}) becomes
\begin{equation}
\delta_k=\frac{1}{N_{\rm p}} \sum_j \sum_{i|J(i)=j}  M_i \exp\{ I
k\cdot [j+\Delta(i)] \},
\end{equation}
with
\begin{equation}
\Delta(i)\equiv x_i-J(i).
\end{equation}
The quantity $\Delta(i)$ 
(or each coordinate of it in more than 1D) is bounded in
$[-1/2, 1/2 [$. To simplify the
expressions, it is assumed without any loss of generality that the size of
a cell of the grid is unity, $\Delta_{\rm g} \equiv 1$.
The Fourier-Taylor expansion of order $N$ is then
\begin{equation}
\delta_k^{(N)} = \frac{1}{N_{\rm p}} \sum_j  \exp( I k\cdot j ) \sum_{n=0}^N
\frac{1}{n!}\sum_{i|J(i)=j} M_i \times [I k\cdot \Delta(i)]^n.
\label{eq:fourtay}
\end{equation}
Such an expansion is expected to converge very quickly as a consequence
of  properties of trigonometric functions, $\sin(x)$ and $\cos(x)$.

With $D$ the number of dimensions, and the vector $\Delta=(\Delta_1,\cdots,\Delta_D)$, from the multinomial theorem,
\begin{eqnarray}
   (k\cdot \Delta)^n & = & \sum_{q_1+\cdots+q_D=n} \frac{n!}{q_1!\times \cdots
  \times q_D!}\times \nonumber \\
  & & \times (k_1 \Delta_1)^{q_1}\times \cdots \times(k_D \Delta_D)^{q_D}.
\end{eqnarray}
Equation (\ref{eq:fourtay}) can thus be rewritten
\begin{eqnarray}
\delta_k^{(N)} & = & \frac{1}{N_{\rm p}} \sum_{n=0}^N I^n
\sum_{q_1+\cdots+q_D=n}  \frac{1}{q_1!\times \cdots \times q_D!}
\times \nonumber \\ & & \times k_1^{q_1}
\times \cdots \times k_D^{q_D} F(k,q) 
\label{eq:resum}
\end{eqnarray}
with
\begin{equation}
F(k,q)={\rm FT}[\mu_q]\equiv \sum_j \mu_q(j) \exp[ I k\cdot j ],
\label{eq:Fouriert}
\end{equation}
where ${\rm FT}$ is the Fourier operator and 
\begin{equation}
\mu_q(j)=\sum_{i|J(i)=j} M_i \times [\Delta_1(i)]^{q_1} \times \cdots
\times [\Delta_D(i)]^{q_D}.
\label{eq:moment}
\end{equation}
This defines the Fourier-Taylor algorithm: the approximate
direct Fourier transform is reduced to 
\begin{enumerate}
\item the  calculation of the moments $\mu_q(j)$ on the the real space
  grid, equation~(\ref{eq:moment});
\item their Fourier transform, equation~(\ref{eq:Fouriert}), which can be
  performed with usual FFT algorithms;
\item their summation with the appropriate weights,
  equation~(\ref{eq:resum}).
\end{enumerate}
Note importantly that periodicity was not assumed in this calculation, and that the
values of (each coordinate of) $k$  available are theoretically any
multiples of $2 \pi/N_{\rm g}$, as a simple consequence  of the periodicity
of  the function $F(k,q)$ in $k$ space. However, the accuracy of the
Taylor expansion is controlled by the magnitude of $k\cdot \Delta(i)$,
and thus worsens with larger $k$. As a result, 
we shall restrict at present time to the naturally available range
of values of (each coordinate of) $k$, $[-k_{\rm ny},k_{\rm ny}]$,
where $k_{\rm ny} \equiv \pi$ corresponds to the Nyquist
frequency defined by the \emph{grid}, which is \emph{a priori} unrelated to the distribution of points. We shall explain in \S~\ref{sec:extension} how to extend
the algorithm to have access to arbitrary values of $k$, 
while maintaining the errors on the Taylor expansion bounded.

It is crucial to point out a few features of this calculation. First, this is very specifically the Fourier transform of a point distribution, which is not precisely equivalent to the transform of an irregularly-sampled continuous function (which requires the further specification of an interpolation scheme). Second, if we restrict the calculation to a finite set of wavenumbers $k$, there is no unique inverse, and we cannot recover the real-space distribution from the Fourier transform. Finally, the lowest-order ($N=0$) version of this calculation is equivalent to nearest-grid-point interpolation to the $N_g^D$ grid.

The algorithm now scales in three dimensions like ${\cal O}[ N_{\rm FFT} \times N_{\rm g}^3 \log
N_{\rm g}] + {\cal O} [ N_{\rm FFT} \times N_{\rm p} ]$, for accessing $N_k
\sim N_{\rm g}^3$ wavenumbers, where $N_{\rm FFT}$ represents
the number of Fourier transforms involved in the calculation. 
If one assumes $N_{\rm p} \ga N_{\rm g}^3$, this method is much faster 
than the direct summation approach if $N_{\rm FFT} \ll N_{\rm g}^3$. 
The parameter $N_{\rm FFT}$ is given by
\begin{equation}
N_{\rm FFT} = \sum_{n=0}^{N} \sum_{q_1+\cdots+q_D=n} 1 =
\frac{(n+D)!}{D!\ n!}.
\end{equation}
Table~\ref{tab:table_num_FFT} gives the corresponding numbers for
$D=1, 2$ and 3.
\begin{table}
\caption[]{Number of Fourier transforms, $N_{\rm FFT}$ and the rough estimate
 of the error, $E(D,N)$, in the worse case (at the Nyquist frequency) according to equation~(\ref{eq:error}), as
 functions of the considered order $N$ of the Fourier-Taylor expansion and the
  number of dimensions $D$ of the system.}
\begin{tabular}{rrrrrrr}
\hline
     & \multicolumn{2}{c}{$D=1$}  & \multicolumn{2}{c}{$D=2$} & \multicolumn{2}{c}{$D=3$} \\
 $N$ &  $N_{\rm FFT}$ & $E(1,N)$ & $N_{\rm FFT}$ & $E(2,N)$ & $N_{\rm FFT}$ & $E(3,N)$ \\
\hline
0  & 1  & 1.6     & 1   & 3.1     & 1    & 4.7    \\
1  & 2  & 1.2     & 3   & 4.9     & 4    & 11     \\
2  & 3  & 0.6     & 6   & 5.2     & 10   & 17     \\
3  & 4  & 0.25    & 10  & 4.1     & 20   & 21     \\
4  & 5  & 0.08    & 15  & 2.6     & 35   & 19     \\
5  & 6  & 2.1   $10^{-2}$  & 21  & 1.3     & 56   & 15     \\
10 & 11 & 3.6   $10^{-6}$  & 66  & 7.4   $10^{-3 }$ & 286  & 0.6    \\
15 & 16 & 6.6   $10^{-11}$ & 136 & 4.3   $10^{-6 }$ & 816  & 2.8   $10^{-3}$ \\ 
20 & 21 & 2.5   $10^{-16}$ & 231 & 5.4   $10^{-10}$ & 1771 & 2.7   $10^{-6}$ \\
\hline
\end{tabular}
\label{tab:table_num_FFT}
\end{table}
The accuracy of the approximation is dictated by the
magnitude of the next order correction in equation~(\ref{eq:fourtay}), 
$[k\cdot \Delta(i)]^{N+1}/(N+1)!$. Errors become more significant at the Nyquist
frequency of the grid and for $\Delta(i) \sim 1/2$. At first sight, control
of the error is thus given by the condition
\begin{equation}
E(D,N)=(\pi D/2)^{N+1}/(N+1)! \la \epsilon,
 \label{eq:error}
\end{equation}
where $\epsilon$ is a small parameter, but of course the actual error
depends on the spectral properties of the system considered. While order $N=10$ is enough
to have $\epsilon < 10^{-5}$ for $D=1$, we need $N=15$ and $N=20$ for
$D=2$ and $D=3$, respectively. For this  level of accuracy, the computational cost becomes
increasingly prohibitive for increasing  value of $D$ due to the large number of
Fourier transforms required to perform the calculations. This makes the Fourier-Taylor approximation
mainly attractive for $D=1$ if one aims to estimate $\delta_k$
accurately for any value of $k$. However, the goal here is not to have
an accurate measurement of $\delta_k$ but rather of its
power spectrum. Let us now  investigate how the Fourier-Taylor
method behaves for this latter quantity.

\section{The Fourier-Taylor power spectrum}
\label{sec:ftps}
This section is divided into two parts. In \S~\ref{sec:powspecesti}, 
we compute the ensemble
average of the naively-defined rough Fourier-Taylor power spectrum, assuming that the point process
under consideration is a local Poisson realization of a stationary
random field, periodic over the grid used to run the
Fourier-Taylor algorithm. For instance we shall recover a well-known
result for nearest grid point interpolation (NGP), which corresponds
to the zeroth-order Taylor expansion. 
In \S~\ref{sec:biasplusfolding}, we analyse the various biases in the rough Fourier-Taylor
estimator, namely the shot noise of the particles which
can be subtracted off, the bias due to the
interpolation method (\emph{e.g.}, the famous ${\rm sinc}^2$ biasing from NGP
interpolation), which can be easily corrected for, and the residual
due to aliasing. The calculations here do not necessarily
assume isotropy of the underlying random process, but the 3D analyses
use angular averages, which make sense only if isotropy applies.

\subsection{Ensemble average for a stationary point process}
\label{sec:powspecesti}
In what follows, we assume that the catalog is a set of particles of
equal weights, $M_i=1$. A naive estimate of the power spectrum of order
$N$ can be written
\begin{equation}
P^{(N)}(k)\equiv \frac{1}{{\bar N}_{\rm p}^2}\left\langle   N_{\rm p}^2\ \delta_k^{(N)} \delta_{-k}^{(N)} \right\rangle,
\label{eq:estina}
\end{equation}
where
\begin{eqnarray}
  N_{\rm p}^2\ \delta_k^{(N)} \delta_{-k}^{(N)}  =  \sum_{j,j'}
  \exp[ I k\cdot (j-j') ] \sum_{n,m=0}^N \frac{1}{n!\ m!} \times \nonumber \\
  \times \sum_{i|J(i)=j}\ \sum_{i'|J(i')=j'}
  [I k\cdot \Delta(i)]^n\ [-I k\cdot \Delta(i')]^m,
  \label{eq:pows}
\end{eqnarray}
${\bar N}_{\rm p} =\langle N_{\rm p} \rangle$,
and $\langle \cdots \rangle$ stands for the ensemble average over many
realizations. Note thus that in the following calculation, we shall 
allow the number of objects $N_{\rm p}$ in the catalog to
fluctuate. Also, isotropy is not yet assumed here: $k$ is still a
vector in equation~(\ref{eq:estina}). Setting
\begin{equation}
\nu_n(k)\equiv \int_{-1/2}^{1/2} (k\cdot \Delta)^n d^D\Delta,
\label{eq:nunint}
\end{equation}
and
\begin{eqnarray}
   \nu_{n,m}(k,j-j') & \equiv & \int_{-1/2}^{1/2}
   \xi(j-j'+\Delta-\Delta')\times \nonumber \\
   & & \times(k\cdot \Delta)^n\ (k\cdot \Delta')^m \ d^D\Delta\ d^D\Delta',
  \label{eq:nunm}
\end{eqnarray}
where $\xi(x)$ is the two-point correlation function assumed to be invariant by
translation, ensemble averaging  equation~(\ref{eq:pows}) reads
$$
  \left\langle N_{\rm p}^2 \ \delta_k^{(N)} \delta_{-k}^{(N)}
      \right\rangle  = {\bar N}_p^2\ \delta_{\rm D}(k) + \sum_{n,m=0}^N
      \frac{I^{n-m}}{n!\ m!} \left\{ {\bar N}_p\ \nu_{n+m}(k) +\right.
$$
\begin{equation}
      \null \quad \quad \quad \quad \quad + {\bar N}^2 \sum_{j,j'} \exp[ I k\cdot (j-j') ]\ \nu_{n,m}(k,j-j')
      \},
\end{equation}
where $\delta_{\rm D}(k)$ is the Dirac delta function and 
${\bar N}$ is the average number of particles per cell,
\begin{equation}
{\bar N}={\bar N}_{\rm p}/{N_{\rm g}^3}.
\end{equation}
This calculation can be derived quite easily following the microcells formalism of Peebles
(1980), as explained in Appendix~\ref{sec:a1}. 
Let us assume periodicity over the grid and decompose the function $\xi(r)$ in Fourier modes,
\begin{equation}
\xi(r)=\sum_{-\infty \leq l \leq \infty} P(l) \exp(-I\ l\cdot r),
\label{eq:xisum}
\end{equation}
where $P(l)$ is in fact the sought power spectrum.
Notice that the sum (\ref{eq:xisum}) is infinite because the system is not
necessarily band-limited.  
Then,
\begin{eqnarray}
I^{n-m} \sum_{j,j'} \exp[ I k\cdot (j-j') ]\ \nu_{n,m}(k,j-j')= \nonumber \\
\sum_{j,j'} \sum_{-\infty \leq l \leq \infty} P(l)\ \exp[ I (k-l)\cdot (j-j') ] 
\ \kappa_{n,m}(l,k),
\end{eqnarray}
with
\begin{eqnarray}
\kappa_{n,m}(l,k) \equiv  \int_{-1/2}^{1/2}
\exp[-I\ l \cdot(\Delta-\Delta')]\times \nonumber \\ \times (I\ k\cdot \Delta)^n\ (-I\ k\cdot \Delta')^m
d\Delta\ d\Delta'
\end{eqnarray}
The sums over $j$ and $j'$ cancel unless $l=k+2\pi M$, where
$M$ is an arbitrary (vector) integer. Thus
\begin{eqnarray}
I^{n-m} \sum_{j,j'} \exp[ I k\cdot (j-j') ] \nu_{n,m}(k,j-j')= \nonumber \\
N_{\rm g}^6\ \sum_M {P}(k+2\pi M)  \kappa_{n,m}(k+2\pi M,k).
\end{eqnarray}
Notice that 
\begin{equation}
\kappa_{n,m}(k+2\pi M,k)=\kappa_n(k,M)\times \kappa_m(k,M),
\end{equation}
with 
\begin{equation}
\kappa_{n}(k,M)\equiv\int_{-1/2}^{1/2} \exp[-I\ (k+2\pi M).\Delta]\ (I\ k\cdot \Delta)^n d^D
\Delta.
\label{eq:kapan}
\end{equation}
Details of the  calculation of the (real) number $\kappa_{n}(k,M)$ 
are given in Appendix B.
We thus obtain the simple expression
\begin{eqnarray}
  P^{(N)}(k) & = & \delta_{\rm D}(k) + \frac{1}{{\bar N}_{\rm p}} W_N(k) +
      \nonumber \\
      & & + \sum_M P(k+ 2\pi M) \ \Upsilon_N^2(k,M),
\label{eq:finalres}
\end{eqnarray}
with 
\begin{eqnarray}
\Upsilon_N(k,M) & \equiv & \sum_{n=0}^N \frac{\kappa_n(k,M)}{n!}, \\
W_N(k) & \equiv & \sum_{n,m=0}^N \frac{1}{n! m!} \nu_{m+n}(k).
\end{eqnarray}
It is easy to check from equations~(\ref{eq:nunint}) and (\ref{eq:kapan}) that
\begin{equation}
\lim_{N \rightarrow \infty} \Upsilon_N(k,M)=\delta_{\rm D}(M),
\quad \lim_{N \rightarrow \infty} W_N(k)=1, 
\label{eq:convergence}
\end{equation}
as expected: the Fourier-Taylor approximation tends to the exact
solution, $P(k)+1/{\bar N}_{\rm p}$, when the order $N \rightarrow \infty$.

Using the analytical calculations presented in Appendix B, we recover for 
 $N=0$ (corresponding to NGP
interpolation), the well-known result (\emph{e.g.}, Jing, 2005)
\begin{eqnarray}
P^{(0)}(k) & = & \delta_{\rm D}(k) + \frac{1}{{\bar N}_{\rm
  p}} + \sum_M P(k+2\pi M) \times \nonumber \\ & & \times \prod_{q=1,\cdots,D}
	      [\sin(k_q/2)/(k_q/2+\pi M_q)]^2.
\end{eqnarray}
The bias on the power spectrum introduced by the NGP interpolation 
corresponds, in the ensemble average sense, to the
Fourier transform of the square top-hat function modulo
aliasing of the power spectrum in Fourier space, and can be corrected
for straightforwardly in the band-limited case (where only $M=0$
contributes).\footnote{Note as well that for pure white noise, $P(k)=$constant, NGP is unbiased
since $\sum_M 1/(k_q/2+\pi M_q)^2=1/\sin^2(k_q/2)$.
}

\subsection{Analysis of biases and residuals due to aliasing}
\label{sec:biasplusfolding}
\begin{figure}
\centerline{\hbox{
\psfig{file=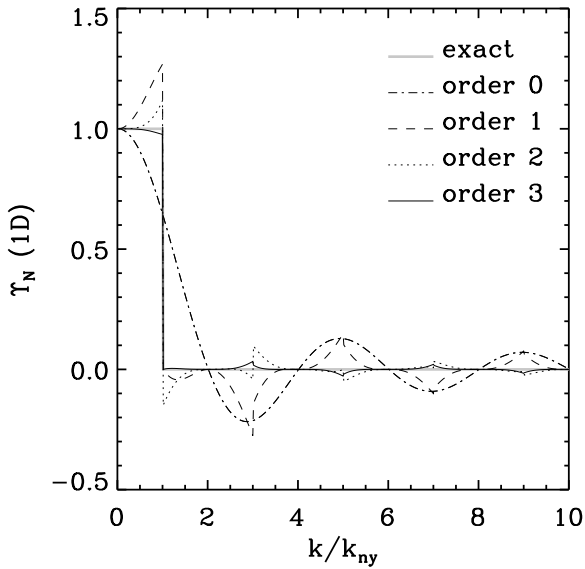,width=8cm}
}}
\caption[]{$\Upsilon_N(k,M)$ as a function of $k/k_{\rm
    ny}+2M$ in the 1D case, for various values of the order $N$ as
    indicated
    on the panel. The calculation has been performed using
   $N_{\rm g}=128$, but the results would not change significantly
   for other values of $N_{\rm g}$. }
\label{fig:1Dfoldings}
\end{figure}

The problem with equation~(\ref{eq:finalres}) is the sum over $M$, as it corresponds
to foldings of Fourier modes at values of $k$ we do not have access
to for a given grid size. The higher the order considered, the smaller the
effect of this aliasing, by construction of the Fourier-Taylor
method. This is illustrated in 1D by Fig.~\ref{fig:1Dfoldings}, which
shows the function $\Upsilon_N(k,M)$
as a function of $k/k_{\rm ny}+2M$, for various orders of the Taylor
expansion. One can see that convergence towards the exact solution
(equation~\ref{eq:convergence}) is rather fast. The effect
of aliasing is largest when approaching Nyquist frequency, a
natural property of the Fourier-Taylor method which is an expansion
around $k=0$.\footnote{Note that a similar polynomial expansion 
(or an expansion on an appropriate basis of functions) to
the Taylor expansion but minimizing
in a global way the effects of the foldings would be more
optimal. One could for instance adapt methods used in
the NFFT algorithm
({\tt www-user.tu-chemnitz.de/$\sim$potts/nfft/doc.php}, 
see, \emph{e.g.}, Potts, Steidl \& Tasche 2001),
potentially more efficient than the 
Fourier-Taylor transform. The advantage of this latter method presented here
is the simple analytic control of all the biases.}
While it is not possible to correct for aliasing of the power spectrum
without additional strong prior assumptions, it is possible to estimate a bound
on the systematic error it induces. For instance, let us assume that
outside the range $[-k_{\rm ny},k_{\rm ny}]$ (for each coordinate
of the wavenumber in more than 1D) the power spectrum
is bounded by a value $P_{\rm max}$:
\begin{equation}
P(k) \leq P_{\rm max}, {\rm outside \ Nyquist\ range}.
\label{eq:prior}
\end{equation}
This obtains, for example, in the common case of a falling spectrum at high wavenumbers.
Then, noticing that (whatever the number of dimensions)
\begin{equation}
\sum_M \Upsilon_N^2(k,M) = W_N(k),
\end{equation}
we obtain the following bound (omitting the additional trivial additional
term at $k=0$)
\begin{equation}
P(k) \leq  \frac{P^{(N)}(k)-W_N(k)/{\bar N}_{\rm p}}{ \Upsilon_N^2(k,0)} \leq
 P(k)+P_{\rm max} R_N(k),
\label{eq:estimator}
\end{equation}
where the positive residual function $R_N(k)$ is given by
\begin{equation}
R_N(k) \equiv  \frac{W_N(k)}{\Upsilon_N^2(k,0)}-1.
\end{equation}
Equation (\ref{eq:estimator}) defines a range for the estimation of the unbiased
power spectrum with the weak assumption given by equation~(\ref{eq:prior}).

The residual $R_N(k)$ estimates the influence of the foldings of the (unknown) power spectrum
at wavenumbers outside the Nyquist domain defined by the grid. It is
expected to decrease rapidly with order $N$, since 
$W_N(k) \simeq \Upsilon_N^2(k,0)$ when $|k|/k_{\rm ny} \ll 1$: at leading order in
$k/k_{\rm ny}$, after simple algebraic calculations, one finds, for
$N \geq 1$,
\begin{eqnarray}
W_N(k) & \simeq & \Upsilon_N^2(k,0) \\
& \simeq & 1-\frac{2 (-1)^{N/2}}{N!(N+2)}
\nu_{N+2}(k), \quad N\ {\rm even} \label{eq:neven} \\
 & \simeq & 1-\frac{2 (-1)^{(N+1)/2}}{(N+1)!} \nu_{N+1}(k), \quad N\
   {\rm odd}. \label{eq:nodd}
\end{eqnarray}
Equation (\ref{eq:neven}) remains valid for $N=0$ only for
$\Upsilon_0^2(k,0)$, while $W_0(k)=1$.

\begin{figure}
\centerline{\hbox{
\psfig{file=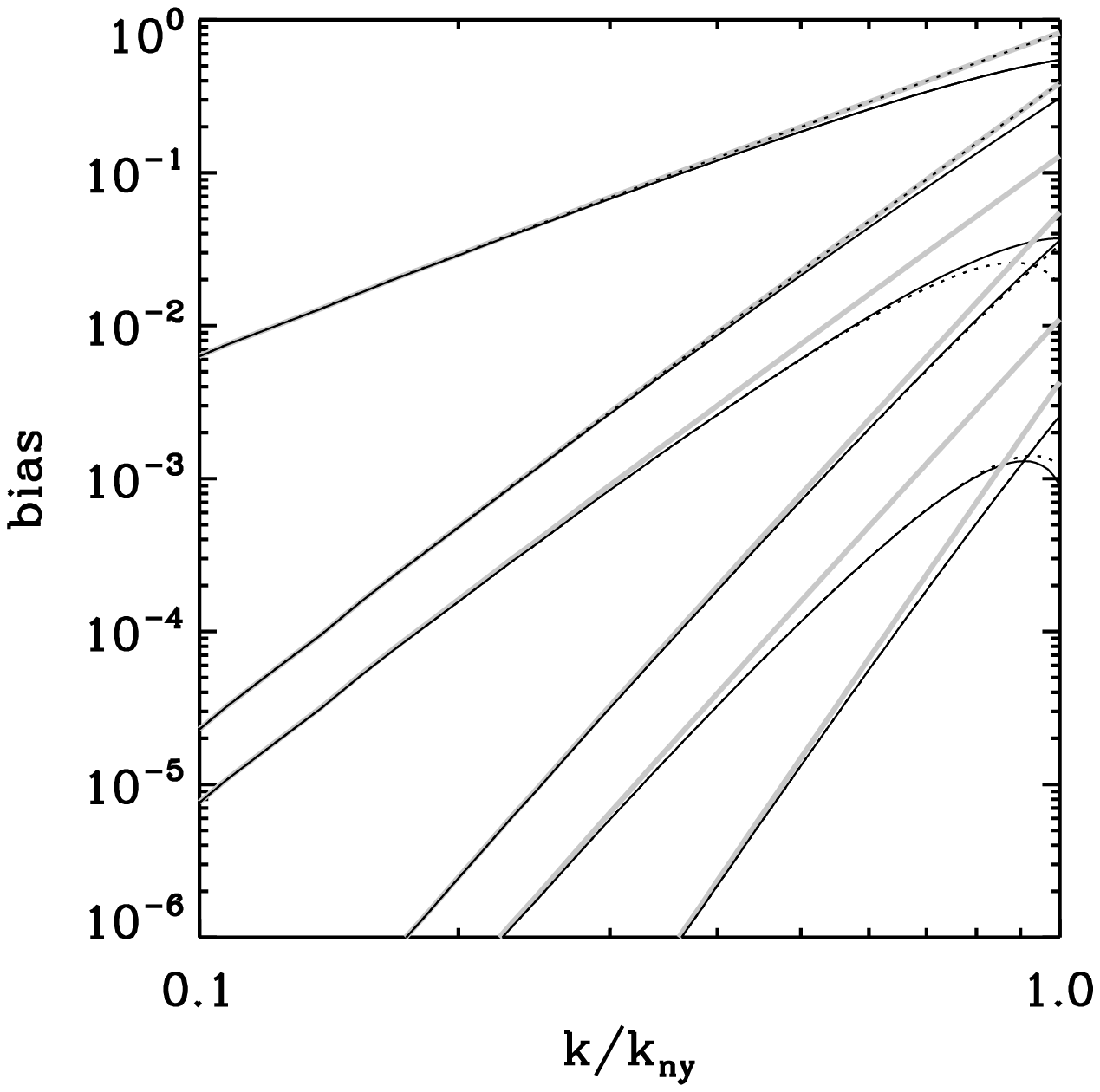,width=8cm}}}
\centerline{\hbox{
\psfig{file=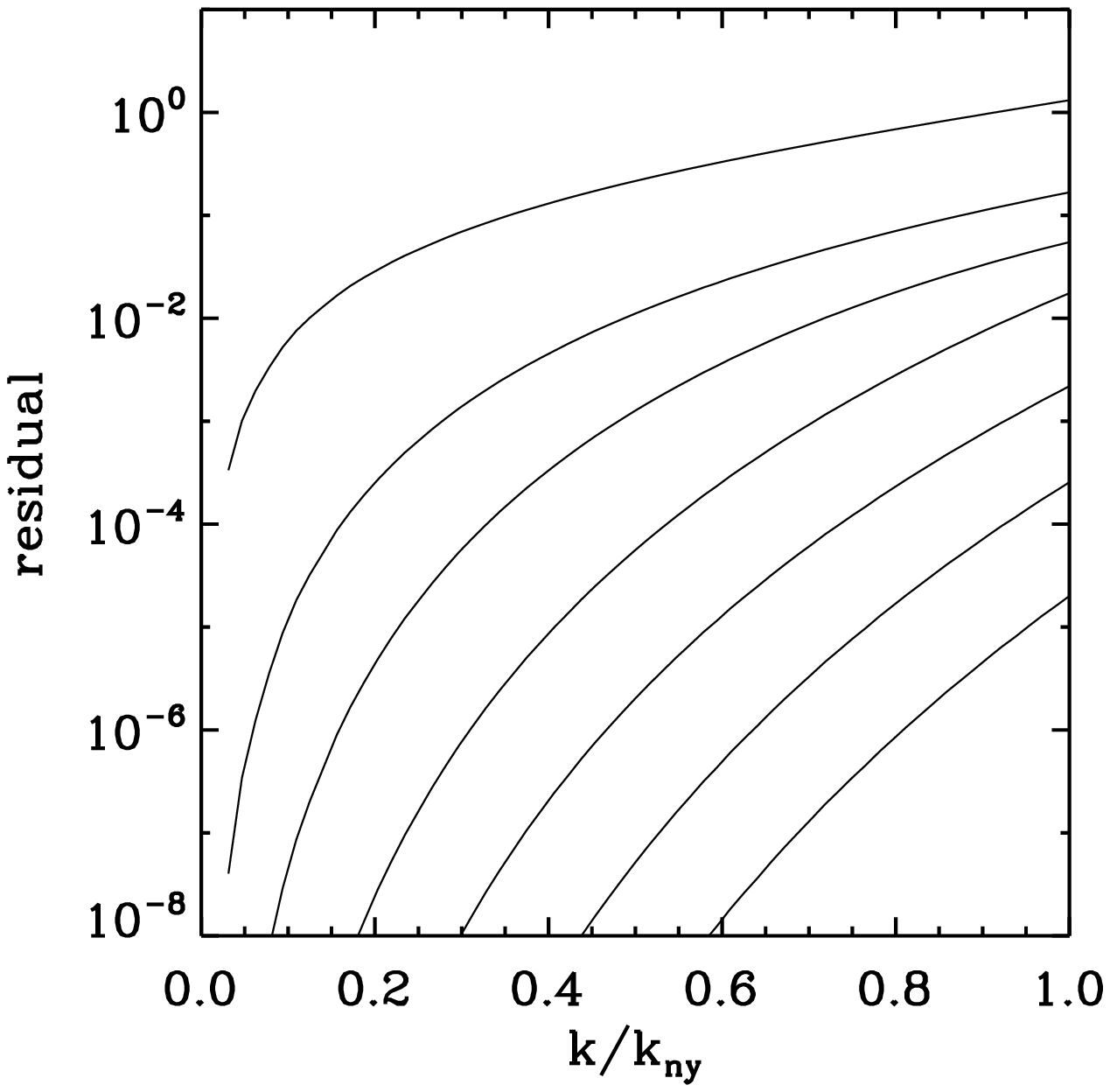,width=8cm}
}}
\caption[]{Biases on the Fourier Taylor spectrum,
$|\Upsilon^2_N(k,0)-1|$, $|W_N(k)-1|$, 
and residual function $R_N(k)$ due to aliasing as functions of $k/k_{\rm ny}$,
after angular average as explained in the main text. The calculation
assumes a grid with $N_{\rm g}=128$, but the results would not change
significantly for other large enough values of $N_{\rm g}$. Each curve corresponds
to a value of the Taylor expansion order, $N$, ranging from $N=0$ to
$N=6$, from top to bottom. On the top panel, the solid, dotted and thick grey curves correspond
to $|\Upsilon^2_N(k,0)-1|$, $|W_N(k)-1|$ and
leading order expression (\ref{eq:nodd}), respectively. The case $N=0$ is omitted,
for clarity. In this latter case, we have $W_0(k)-1=0$
and $|\Upsilon_N^2(k,0)-1|$ is of the same order (but slightly different)
of what is obtained at first order, $N=1$. 
}
\label{fig:residu}
\end{figure}
To illustrate quantitatively these results in the $D=3$ case, 
Figure \ref{fig:residu} shows the angular averages 
of the biases $\Upsilon^2(k,0)-1$, 
$W_N(k)-1$ (left panel) and
the residual $R_N(k)$ as functions of $|k|$ for various
values of $N$. More specifically, and this will be used further for 3D
measurements, one estimates for each integer wavevector $k N_{\rm g}/(2\pi) $ the following
quantity
\begin{equation}
{\tilde k}(k)=E\left( \left|\frac{k N_{\rm g}}{2\pi} \right|+\frac{1}{2} \right),
\end{equation}
where $E(x)$ is the integer part of $x$.
Then, the angular average of quantity ${\tilde A}({\bar k})$ for integer
wavenumber modulus ${\bar k}$ is given by
\begin{equation}
{\tilde A}({\bar k}) \equiv \frac{1}{C({\bar k})} \sum_{k | {\tilde k}(k)={\bar k}}
A(k),
\end{equation}
where the count $C(k)$ is
\begin{equation}
C({\bar k})\equiv \sum_{k | {\tilde k}(k)={\bar k}} 1.
\end{equation}
This gives the number of  integer wavenumbers verifying ${\tilde
  k}(k)={\bar  k}$. Note that angular averages make sense
only if one assumes statistical isotropy, which then means 
that the power spectrum  depends only on $|k|$. This is theoretically
the case of the cosmological random fields considered in this work. 

The way the angular average is performed here is very rough, and itself
introduces some biases with respect to the estimate of the true
angular average of the power spectrum. Implementation of a better
angular averaging procedure would be quite straightforward and easy
to propagate in the analytic calculations, but would not serve
the purpose of this paper, so we leave it for future
work.\footnote{See, \emph{e.g.},  Scoccimarro et al.~(1998), for
a better handling of angular averages.} 

Figure \ref{fig:residu} shows that the residual function $R_N(k)$
decreases rapidly with $N$. It is supplemented with
Table~\ref{table:tableresi}, which provides numerical values for
$k=k_{\rm ny}/2$ and $k=k_{\rm ny}$. One can already see  
the virtue of the Fourier-Taylor method: going to higher order
reduces these otherwise uncontrollable effects, which are clearly not negligible at
all for the traditional NGP method ($N=0$) where the residual is
of order unity at the Nyquist frequency and still about 20 percent
at half the Nyquist frequency, whereas the third-order Fourier-Taylor correction reduces
it to about 2 percent and $6\times 10^{-5}$, respectively.

\begin{table}
\caption[]{Numerical estimate in the 3 dimensional case
   of the residual function $R_N(k)$ after
  angular average as explained in the text. The calculation has been
  performed with $N_{\rm g}=128$ but the results should not change
  significantly for other values of $N_{\rm g}$ as long as they are
  not too small. The first column indicates the order $N$ of the Taylor
  expansion, while the second and the third one give
  $R_N(k_{\rm ny}/2)$ and $R_N(k_{\rm ny})$, respectively.}
\vskip 0.5cm
\begin{tabular}{r|rr}
\hline
  $N$  & $R_N(k_{\rm ny}/2)$ & $R_N(k_{\rm ny})$ \\
\hline
0  & 0.22    & 1.3      \\
1  & 0.011   & 0.17     \\
2  & 1.3   $10^{-3}$  & 0.055    \\
3  & 5.6   $10^{-5} $ & 0.018    \\
4  & 2.0   $10^{-6} $ & 2.2   $10^{-3}$   \\
5  & 5.2   $10^{-8} $ & 2.6   $10^{-4} $  \\
6  & 1.1   $10^{-9} $ & 2.0   $10^{-5}  $ \\
\hline
\end{tabular}
\label{table:tableresi}
\end{table}

To understand better the scaling of the functions $\Upsilon^2_N(k,0)-1$
and $W_N(k)-1$ with $k$, one can perform the integral of
equation~(\ref{eq:nunint}) in a sphere instead of a cube.
For $D=3$, it reads
\begin{equation}
\nu_N(k)\simeq \frac{3}{2}\frac{1-(-1)^{N+1}}{(N+1)(N+3)}  
\left( \frac{3}{4\pi}\right)^{N/3} \left( \frac{\pi |k|}{k_{\rm ny}}\right)^N.
\end{equation}
This approximation is not accurate enough for practical calculations, but
allows one to see that $W_N(k)-1$ and $\Upsilon^2_N(k,0)-1$ scale
as $|k|^{N+2}$ for $N$ and $N+1$, $N$ even. For instance, the
second and third order scale the same way with $|k|$, as can been seen
on top panel of Fig.~\ref{fig:residu}. Note interestingly the bending
of the bias for odd $N$ observed when one approaches the Nyquist
frequency: this follows from the fact that  the sine function cancels at
$k=k_{\rm ny}$, unlike the cosine function. Therefore, if one
wants to use the brute-force Fourier-Taylor expansion without a bias
correction to the power spectrum, it is better to perform it at odd
orders.

\section{Validation: tests on an $N$-body simulation}
\label{sec:validation}
In this section, we validate the analytic results derived just above
by performing measurements in a CDM $N$-body simulation. The
simulation is described in \S~\ref{sec:CDMsam}, where we also explain
how we estimate statistical errors on the measurement of the
power spectrum. In \S~\ref{sec:goodgood}, biases on the Fourier-Taylor rough estimator are
measured and compared to the theoretical predictions in the final stage
of the simulation, which should agree well with the assumptions
of local Poisson sampling of a stationary random field used
in the previous section. In
\S~\ref{sec:badbad}, we consider the initial conditions of the simulation, which
correspond to a slightly perturbed grid and therefore strongly
deviate from local Poisson behavior. Finally, in
\S~\ref{sec:unbiases}, we propose a nearly
unbiased estimator that should work in all the cases, disregarding the aliasing
effects on the power spectrum, which are controlled by the order of the Fourier-Taylor
expansion as well as the ratio of $k/k_{\rm ny}$ where $k_{\rm ny}$
is the Nyquist frequency of the grid used to perform the calculations.

\subsection{The CDM {\tt GADGET} sample}
\label{sec:CDMsam}

\begin{figure*}
\centerline{\hbox{
\psfig{file=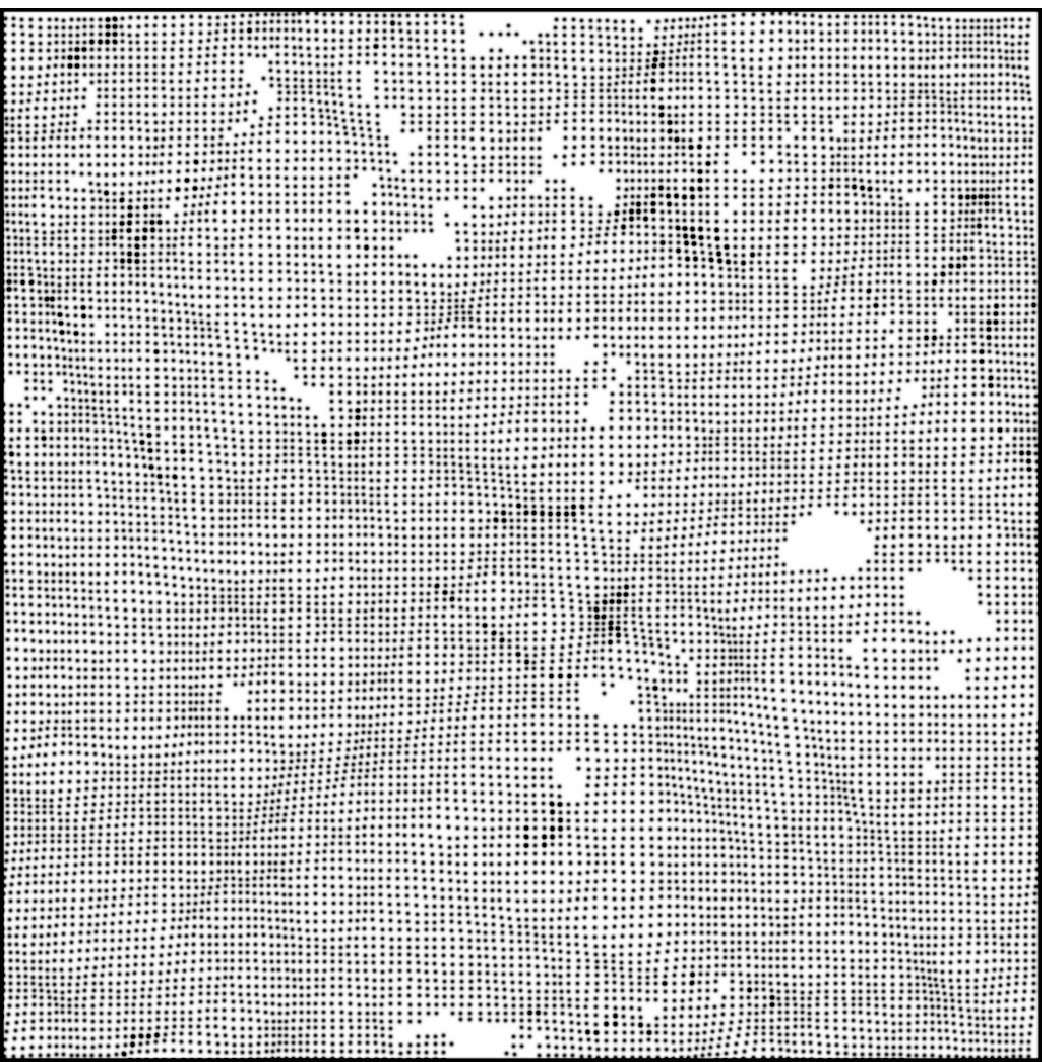,width=8cm}
\psfig{file=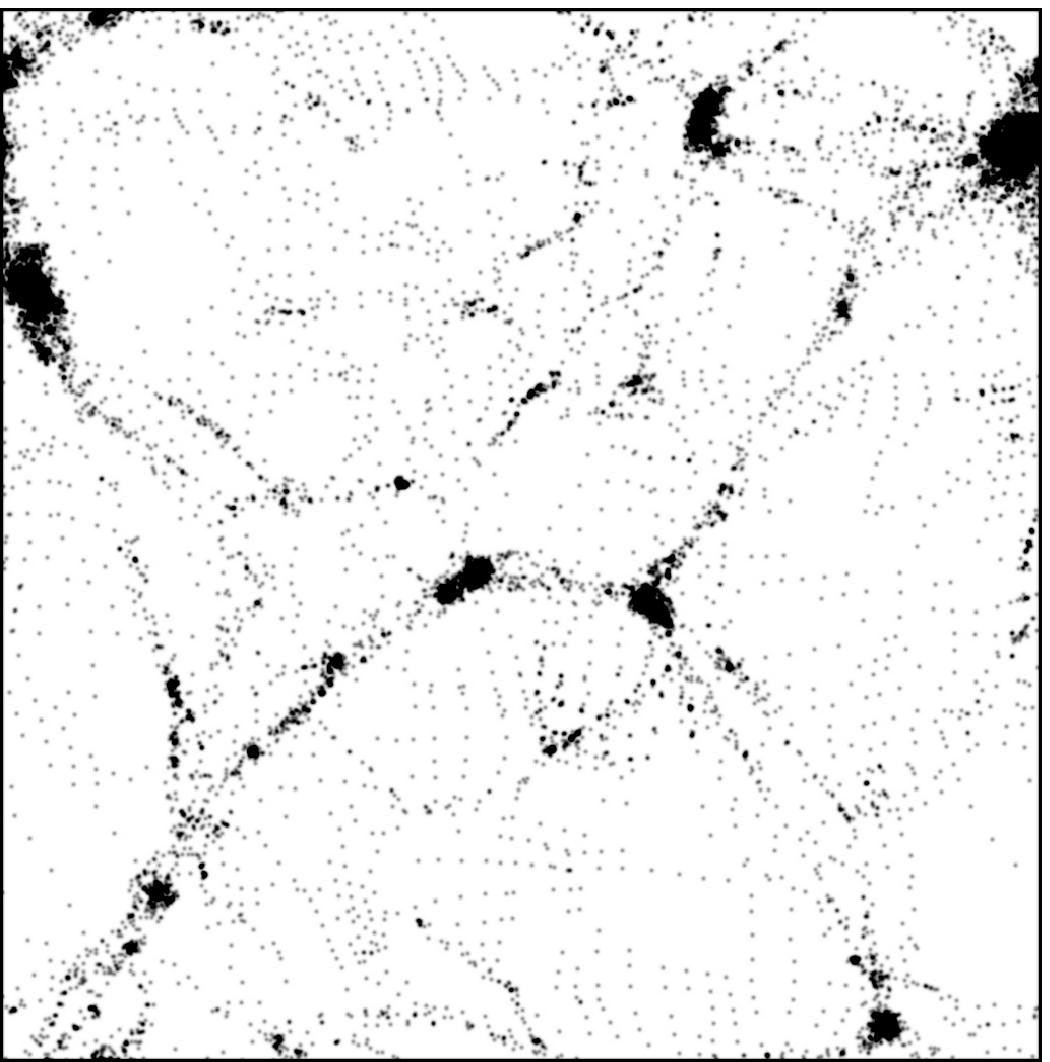,width=8cm}
}}
\caption[]{A thin slice, $L/128$ thick, extracted from the initial
  conditions and the final stage of our $128^3$ CDM {\tt GADGET} simulation.
On the left panel, because of the strong deviations from local Poisson
behavior brought by the grid pattern, the validity of our 
calculations for the power spectrum biases are
questionable. On the right panel, the effects of the grid are much
  less present, although still visible in underdense regions: 
they should have much less impact on the measurements.}
\label{fig:image}
\end{figure*}
We now validate the above results using a Cold Dark
 Matter (CDM) $N$-body simulation performed with the publicly
 available tree code {\tt GADGET} (Springel, Yoshida \& White, 2001), as shown in
Fig.~\ref{fig:image}. The parameters of this simulation are
the following. It uses $N_{\rm p}=128^3$ particles in a periodic box
of size $L=50\ h^{-1}$ Mpc, where $h=H_0/(100\mathrm{km}/\mathrm{s}/\mathrm{Mpc})=0.7$. The cosmological
parameters are matter density $\Omega_{\rm m}=0.3$ and cosmological
constant $\Omega_{\Lambda}=0.7$. The linear variance $\sigma_8^2$ of the density
fluctuations in a sphere of radius $8\ h^{-1}$ Mpc extrapolated
to present time was taken to be $\sigma_8=0.92$. Finally, the
softening length $\varepsilon$ was chosen to be $1/20$th of the mean interparticle
distance. The initial conditions were generated using the {\tt Graphic}
 package of Berstchinger (2001), with a transfer function given by Bardeen et
 al. (1986) (no baryons). This package basically allows one to perturb
an initially homogeneous particle distribution using the Zel'dovich
approximation (Zel'dovich, 1970). For the initial pattern, we chose to put the particles
on a regular grid (see left panel of Fig.~\ref{fig:image}). 
Our test sample is somewhat small compared to
contemporary numerical experiments, but was chosen such that we could
perform ``exact'' calculations of power spectra in a reasonable amount of
 time. By exact, we mean the 20th order Fourier-Taylor expansion using
an $N_{\rm g}=128$ grid. To be able to probe the highly nonlinear 
regime well and to make sure that the system has properly relaxed to a
 locally Poissonian and isotropic stage in collapsed objects (see
right panel of Fig.~\ref{fig:image}), 
we purposely used a small box size. 

\begin{figure}
\centerline{\hbox{
\psfig{file=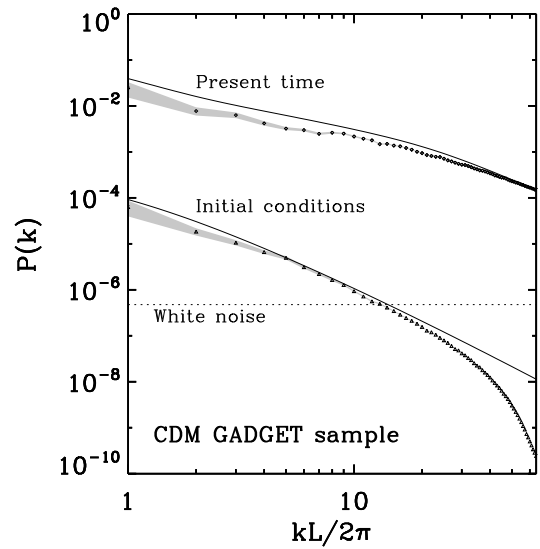,width=8cm}
}}
\caption[]{The power spectrum measured in the initial conditions
  (lower curves) and the last snapshot, corresponding
to the present time (upper curves), of our $128^3$
  CDM {\tt GADGET} sample. The symbols correspond
to the ``exact'' measurement with the Fourier-Taylor method of order 20.
A shaded region is superposed on them. This represents uncertainties
on the measurements as computed from equation~(\ref{eq:errorint}).
The smooth curve gives the non-linear ansatz of Peacock \& Dodds (1996).
The horizontal dotted line corresponds to the white noise level of
the particle distribution. Note that
for the measurement at the present time, a correction for white noise was
  performed, but not for the initial conditions measurement. The
  bending
of the power spectrum at high $k$ in the latter case is due to the
Hanning filtering performed in the {\tt Graphic} package of Berstchinger (2001).}
\label{fig:powspec}
\end{figure}
The ``exact'' power spectrum of the particle distribution in this sample 
is shown in Fig.~\ref{fig:powspec}, both for the  initial conditions and the final stage of
the simulation.
The smooth curves correspond to the nonlinear ansatz of Hamilton et
 al.~(1991) using the formula of Peacock \& Dodds (1996), shown
here for reference. The error bars represented by the gray shadded
areas correspond to the following self-consistent estimate of the
statistical error:
\begin{eqnarray}
E^2({\bar k}) & = & \left( \frac{\Delta {\tilde P}_{\rm rough}({\bar k})}{{\tilde
    P}_{\rm rough}({\bar k})} \right)^2 \nonumber \\
 &= & \frac{1}{C({\bar k})[C({\bar k})-1]} \times \nonumber \\
 & & \times \left[ \sum_{k/{\tilde
    k}(k)={\bar k}} (\delta_k \delta_{-k})^2 -
     C({\bar k}) {\tilde P}^2_{\rm rough}({\bar k}) \right],
\label{eq:errorint}
\end{eqnarray}
where 
\begin{equation}
{\tilde P}_{\rm rough}({\bar k})=\frac{1}{C({\bar k})} \sum_{k/{\tilde
    k}(k)={\bar k}} \delta_k \delta_{-k}
\end{equation}
is the rough power spectrum measured from the distribution  of particles.
From now on we omit the cumbersome tilde on ${\tilde P}$ and bar on ${\bar k}$
as we shall only consider angular averages until the end of this section.
In the framework of an isotropic, stationary local Poisson process
discussed in \S~\ref{sec:ftps},
recall that after ensemble averaging over many realizations
\begin{equation}
\langle P_{\rm rough}(k) \rangle =   P(k) + \frac{1}{N_{\rm p}},
\end{equation}
where $P(k)$ is the underlying power spectrum. 
We noticed that the error given by equation~(\ref{eq:errorint}) 
is well-approximated by the well-known 
result obtained for a random Gaussian field (\emph{e.g.}, Feldman, Kaiser \&
Peacock, 1994)
\begin{equation}
E^2_{\rm G}(k)=\frac{1}{C(k)},
\label{eq:errgaus}
\end{equation}
which translates, for the desired shot-noise-corrected power spectrum to
\begin{equation}
\left( \frac{\Delta P}{P} \right)^2=\frac{1}{C(k)} \left[ 1 + \frac{2}{N_{\rm
    p} P(k)} + \frac{1}{N_{\rm p}^2 P^2(k)} \right].
\label{eq:errPoisson}
\end{equation}
Here,  $C(k)$ represents the number of \emph{statistically-independent} available wavenumbers, hence the missing factor of two compared to the usual case, since
symmetries in $k$-space are already taken into account. We are a bit
puzzled by the very good agreement between equation~(\ref{eq:errorint}) and
(\ref{eq:errgaus}), as we would expect non-Gaussian
contributions to the error in equation~(\ref{eq:errorint}) due to
nonlinear coupling generated by the dynamics. It seems that these couplings are rather small, as already noticed by Rimes
\& Hamilton (2006) and Hamilton, Rimes \& Scoccimarro (2006). 
Of course, we know for
sure that the error given by equation~(\ref{eq:errorint}) and
(\ref{eq:errgaus}) underestimate the true value in general, that
would be obtained from the dipersion over many realizations of our
simulation (Scoccimarro, Zaldarriaga \& Hui, 1999). 
However, the realistic calculation of such an error requires
prior knowledge of the bispectrum and the trispectrum. Moreover, eqs.~(\ref{eq:errorint}) and
(\ref{eq:errgaus}) are sufficient to prove the points discussed
in the analyses of this paper and to provide a rough estimate of errors in a simple and
self-consistent way, as can be provided by the numerical package we propose.\footnote{For further discussion of statistical errors on the power spectrum,
in particular some possible improvements of equation~(\ref{eq:errorint})
for a self-consistent calculation of the errors and 
the covariance matrix of the measured power spectrum, see Hamilton,
Rimes \& Scoccimarro (2006).}

Due to the very small size of the box, 
the agreement between the smooth curves and the measurement on Fig.~\ref{fig:powspec}
is not very good at large scales (small $k$), where few individual
modes are available; this effect is even worse
at the final stage of the simulation because of  the nonlinear coupling at scales
close to the simulation box size. Note importantly
that the choice of a regular pattern combined with
the Zel'dovich approximation for the initial particle distribution
has a non-trivial influence on the evolution of individual
modes (Marcos et al., 2006; Joyce \& Marcos, 2007a,b; 
see also Crocce, Pueblas \& Scoccimarro, 2006, who discuss transients
coming from using the Zel'dovich approximation). 
Note also that the damping of the power spectrum at large $k$ measured
in the initial conditions is not due to any interpolation effect --- as
we have access here to an ``exact'' measurement --- but 
to the Hanning filtering performed in {\tt Graphic} (see Bertschinger,
2001). 
Finally, while a
shot-noise correction was performed on the $P(k)$ obtained 
in the final stage of the simulation, \emph{i.e.}, a term $1/N_{\rm p}$ was subtracted 
from the rough measurement, we reiterate that
it does not apply to the initial stage. In this case, it is more
appropriate to perform no correction as we are in the situation
of a perturbed grid pattern (\emph{e.g.}, Joyce \& Marcos,
2007a).\footnote{This can be easily checked by analysing
equation~(\ref{eq:powergrid}) which gives the power spectrum of a perturbed
grid using the Zel'dovich approximation.} 

\subsection{The ideal situation: final, relaxed stage}
\label{sec:goodgood}
\begin{figure*}
\centerline{\hbox{
\psfig{file=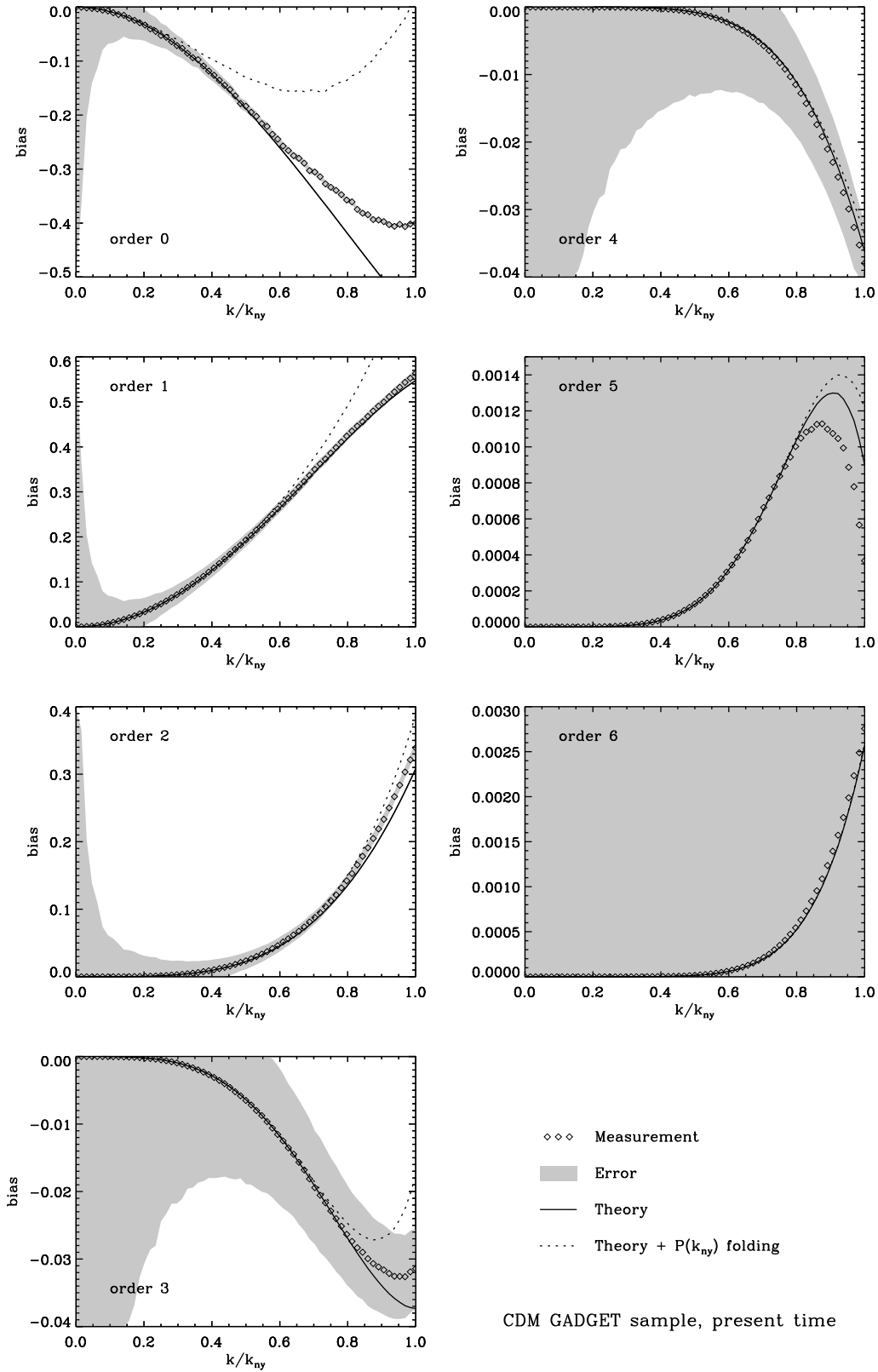,width=14cm}
}}
\caption[]{The measured bias, $b(k)$ (equation~\ref{eq:theobound}), 
on the rough estimator of the power spectrum, $P^{(N)}(k)$, as a
function of $k$ for our CDM {\tt GADGET} sample (symbols). 
Each panel corresponds to a value of the order
$N$, as indicated.
The solid and dotted curves represent a theoretical
lower and upper bound, respectively, in
between which the symbols should lie 
[equation~\ref{eq:theobound} with $P_{\rm max}=P(k_{\rm ny})$], in the
framework of \S~\ref{sec:ftps} ---
local Poisson sampling of an isotropic stationary random
field.  The shaded region represents the statistical error computed from
equation~(\ref{eq:errorint}).}
\label{fig:simuz0}
\end{figure*}
Figure \ref{fig:simuz0} shows the measured bias on the shot-noise-corrected measured Fourier-Taylor power spectrum of order $N$, for
various values of $N$:
\begin{equation}
b(k)=\frac{P^{(N)}(k)-W_N(k)/N_{\rm p}-P(k)}{P(k)}.
\end{equation}
If the analytic calculations of \S~\ref{sec:biasplusfolding} apply, we should have
\begin{equation}
\gamma_N^2(k,0)  \la b(k) \la \gamma_N^2(k,0)  + \frac{P_{\rm max}}{P(k)}
  [ W_N(k)-\Upsilon_N^2(k,0)].
\label{eq:theobound}
\end{equation}
The lower and upper bound are represented by the continuous and the
dotted line respectively. We assume $P_{\rm max}=P(k_{\rm ny})$.
As expected, within the statistical errors defined by equation~(\ref{eq:errorint}),
the measurements represented by the symbols indeed lie in between these two
curves. Since the true
power spectrum is a strongly decreasing function of $k$ in the CDM
cosmology, the effects of aliasing of the power spectrum are
overestimated by the dotted curve, so the symbols are much closer 
to the solid curve than to the
dotted one. In fact, they overlay quite well on the solid curve
when $k$ is small enough compared to $k_{\rm ny}$, as expected.
Note that the nearest grid point interpolation (upper left panel) is still
very significantly contaminated by aliasing, while 
this effect decreases rapidly with the order $N$ as shown in previous
section. The bias also becomes smaller and smaller with $N$, but it is really worth
correcting for, since  the theoretical predictions match rather well the
measurements. Note that at high order, $N \geq 4$, the symbols
no longer lie between the dotted and solid curve, but this is nonetheless
well within the statistical errors represented as the shaded region. 
Moreover, the system
still deviates locally from a pure random stationary pattern (right panel of
Fig.~\ref{fig:image}) so one cannot expect the theoretical bound
given by equation~(\ref{eq:theobound}) to remain valid 
at such a level of accuracy.

\subsection{A poor situation: a perturbed grid as in the
  initial stage}
\label{sec:badbad}
\begin{figure*}
\centerline{\hbox{
\psfig{file=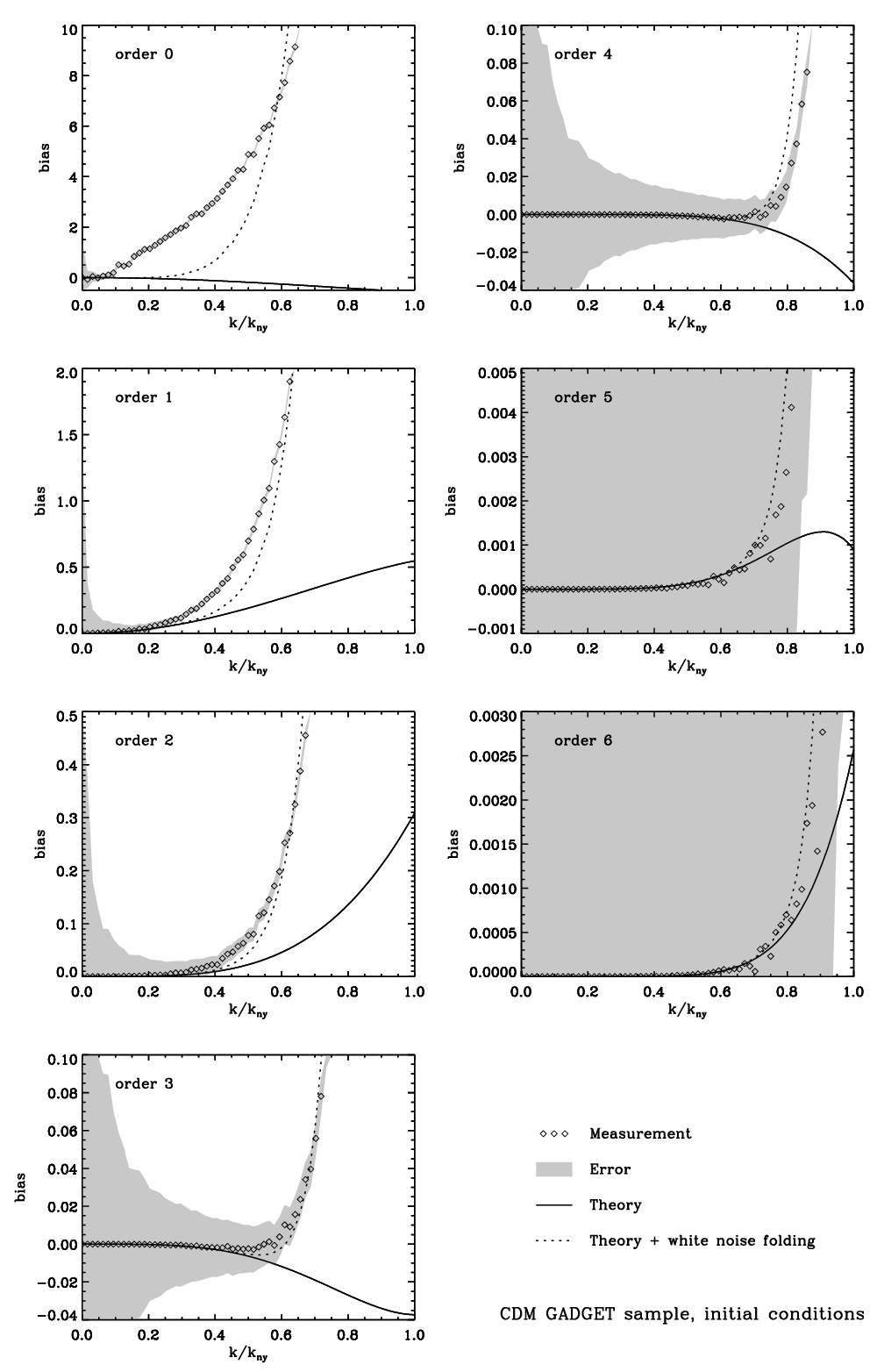,width=14cm}
}}
\caption[]{Same as Fig.~\ref{fig:simuz0}, but the function displayed
is now $g(k)$ (equation~\ref{eq:gdek}) as measured in the initial
conditions of our {\tt GADGET} sample, which correspond to a perturbed
grid. The solid lines on each panel still show function
$\gamma^2_N(k,0)$, while the dotted line gives
$\gamma^2_N(k,0) + (P_{\rm rough}(k) N_{\rm p})^{-1} [W_N(k)-\Upsilon_N^2(k,0)]$ as discussed
in the text.}
\label{fig:simuini}
\end{figure*}
While the final stage of our {\tt GADGET} sample is well within the
framework of the assumptions of local Poisson sampling of a stationary
random process, this is not the case for the initial conditions, which
correspond to a slightly perturbed grid pattern, as shown on left
panel of Fig.~\ref{fig:image}. Fig.~\ref{fig:simuini} is the same
as Fig.~\ref{fig:simuz0}, but now the function measured is
\begin{equation}
g(k)=\frac{P^{(N)}(k)-P_{\rm rough}(k)}{P_{\rm rough}(k)}.
\label{eq:gdek}
\end{equation}
In other words, no correction for the shot-noise of the particles is
performed since it does not make sense in that case.

Before pushing  this analysis further, we have to understand
what is the particle sample at hand.
The {\tt Graphic} package perturbs an initial grid pattern
with a Gaussian random displacement field, ${\cal P}(q)$, supposed to be stationary,
isotropic and curl-free. We assume, to simplify
the discussion which follows, that the grid of particles is the same as the grid used to perform
fast Fourier transforms, $N_{\rm p}=N_{\rm g}^3$. This is the case for
Fig.~\ref{fig:simuini}. 
Remember finally that lengths are expressed in units of the size of the a cell of the grid.

The perturbed position of the particle
is given by
\begin{equation}
x=s+q+{\cal P}(q),
\end{equation}
where $q$ is an integer vector for which each coordinate
is in the range $[0,N_{\rm g}-1]$, and $s$ is
a small constant offset, for which each coordinate is in $[0,1[$. 
Usually, $s=0$ or  $s=(0.5,0.5,0.5)$. 
The Fourier mode of the perturbed grid pattern is given by
\begin{equation}
\delta_k=\frac{1}{N_{\rm p}} \sum_q \exp\{ I k\cdot [s+q+{\cal P}(q)]\}.
\label{eq:Fouriergrid}
\end{equation}
While it is possible to compute the exact power spectrum of such a perturbed grid
(Gabrielli, 2004, see Appendix C and equation~\ref{eq:powergrid} below), 
the general calculation for the Fourier-Taylor expansion is
very cumbersome. However,  three interesting regimes can at least partly be
discussed qualitatively:
\begin{enumerate}
\item {\em Very small displacement, $\sigma^2 \equiv \langle {\cal P}^2
  \rangle \ll 1$:} in that case, the value of $s$ controls the
   amplitude of the displacement $\Delta(i)$ in
   equation~(\ref{eq:fourtay}) and therefore the accuracy of the
  Fourier-Taylor expansion. The smaller  $s$, the better.
  The worst situation is when $s$ approaches diagonal values, 
  \emph{e.g.}, $s= (0.5,0.5,0.5)$. Therefore, to best
 exploit the Fourier-Taylor method on a very slightly perturbed grid,
 it is wise to chose  $s$  (and more generally the size of the
 grid $N_{\rm g}$) carefully, to minimize as much as possible the amplitude
 of the displacements $\Delta(i)$. In the conventions used here,
the wisest choice is thus $s=0$. 
\item {\em Significant displacement, $\sigma^2 \simeq 1$:}
this is the situation of Fig.~\ref{fig:simuini}, where $\sigma \simeq
0.88$. In that case, there
will be always a large fraction of particles with large magnitude of the
displacement $\Delta(i)$, independently of the choice of the offset
$s$. 
\item {\em Large displacement, $\sigma^2 \gg 1$:}
  it is  difficult in this case
  to make any quantitative statements because the conclusions depend
  on the coherence of the displacement field
  (see equation~\ref{eq:powergrid} below). However, one can
  postulate in general that the information due to the grid pattern has
  become subdominant: the particle 
  distribution should behave again like the local 
  Poisson realization of an
  isotropic stationary random process and the calculations of
  \S~\ref{sec:ftps} should in practice become valid
  again.
\end{enumerate}
From this simple discussion, which could be easily extended
to the more general case $N_{\rm p} \neq N_{\rm g}^{3}$, 
one sees that the  measurement of
the power spectrum on a perturbed grid has to be performed carefully in
order to reduce as much as possible the systematic errors on the
Fourier-Taylor spectrum. However, one expects these errors to become
significant only when approaching the Nyquist frequency.

This is well illustrated by  Fig.~\ref{fig:simuini}, for which we
have $s=(0,0,0)$ and a mean square displacement of order 
unity [point (ii) above]: the magnitude of function $g(k)$ increases quite rapidly
when  $k/k_{\rm ny}$ approaches unity. 
The solid line on each panel still represents the
function $\gamma_N^2(k,0)$. The dotted line corresponds to
the right member of equation~(\ref{eq:theobound}), but with
$P_{\rm max}=1/N_{\rm p}$.\footnote{and of course with $P(k)$ replaced
by $P_{\rm rough}(k)$ (with no shot noise correction).} 
This gives a good idea of the overall
behavior of $g(k)$, and this is not very surprising.
Indeed the calculations of Appendix C (see also, \emph{e.g.}, Gabrielli,
2004) give \\

\noindent $
\langle \delta_k \delta_{-k} \rangle  =  \delta_{\rm D}(0)\ +
$
\begin{equation}
  \null \quad \quad +\frac{1}{N_{\rm
    p}} \sum_{q} \exp( I k  \cdot q )
    \exp\left\{ -\frac{k^2 \sigma^2}{3} \left[ 1-\rho(q) \right] \right\},
    \label{eq:powergrid}
\end{equation}
where  $-1 \leq \rho(q) \leq 1$ is the correlation coefficient of
the displacement field.
From this equation, one sees that for a moderately perturbed grid ($\sigma^2 \la 1$), the power spectrum
presents a peak at twice the Nyquist frequency of the grid. 
The amplitude of this peak is controlled by the amplitude of
the displacements, \emph{i.e.}, the value of $\sigma$. 
However, when $k^2 \gg 1/\sigma^2$, the sum in
equation~(\ref{eq:powergrid}) is dominated by the $q=0$ term, hence
\begin{equation}
\langle \delta_k \delta_{-k} \rangle \simeq \frac{1}{N_{\rm p}},\quad k^2 \gg 1/\sigma^2.
\end{equation}
This means that at wavenumbers large enough, the rough power spectrum
of the perturbed grid is dominated by the shot noise of the
particles. Since $\sigma \sim 1$ in our experiment, this should
happen for $k^2$ approaching a few units (except for
the peak just mentioned above at twice the Nyquist frequency of the
grid). As a result, the effects of the aliasing of the
power spectrum should be roughly of the order  
of the shot noise,
which is indeed the case on Fig.~\ref{fig:simuini}:  the
symbols follow roughly the dotted curve, except perhaps for the nearest
grid point case (upper left panel).

Despite the much stronger effects of aliasing on the Fourier-Taylor
spectrum for the perturbed grid than for the locally Poissonian case, the
systematic errors on $P^{(N)}(k)$ still decrease rapidly with the order, $N$.
They remain quite moderate when $k/k_{\rm ny}$ is kept small enough,
for instance $k/k_{\rm ny} \la 1/2$ for third order, $N=3$, and
increase rapidly if $k$ approaches the Nyquist frequency.

\subsection{Unbiased estimator: a zoom in the range $[0,k_{\rm ny}/2]$}
\label{sec:unbiases}
From equation~(\ref{eq:estimator}) we can write an angle-averaged estimator of the true
power spectrum:
\begin{equation}
P_{\rm  est}^{(N)}(k)=\left\langle
\frac{P^{(N)}(k)}{\gamma^2_N(k,0)}-\frac{R_N(k)+1}{N_{\rm p}} \right\rangle_{\rm angles}.
\label{eq:estimator2}
\end{equation}
Recall that the second term of the right side of this equation has to be ignored if we
consider a perturbed regular pattern such as initial conditions of a
$N$-body simulation. We explicitly write the angular average in this
expression, to show how it should be performed to get the best
estimate of $P(k)$. Equation (\ref{eq:estimator2}) gives the
estimator we propose in that paper. 

\begin{figure}
\centerline{\hbox{
\psfig{file=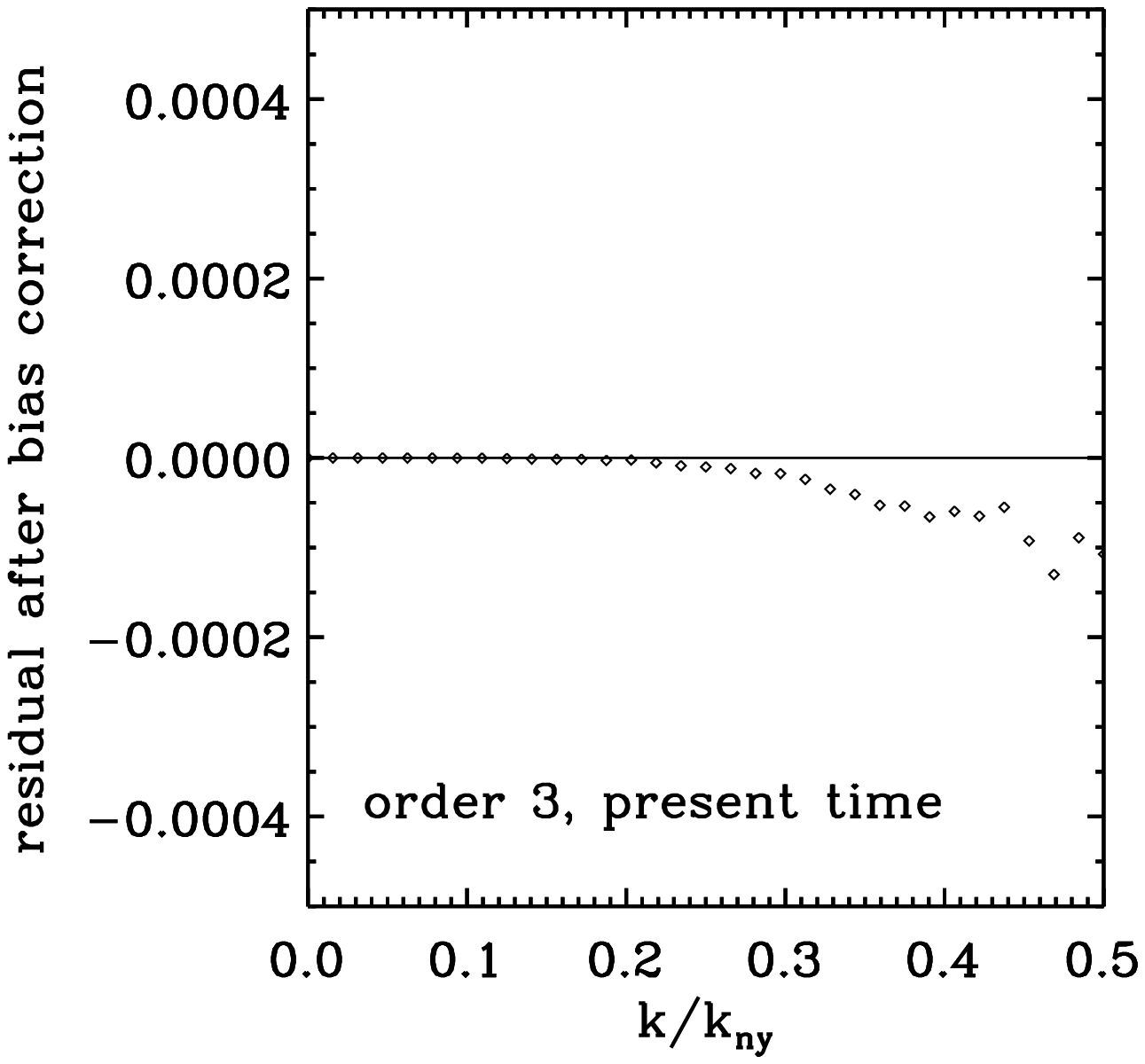,width=8cm}}}
\centerline{\hbox{
\psfig{file=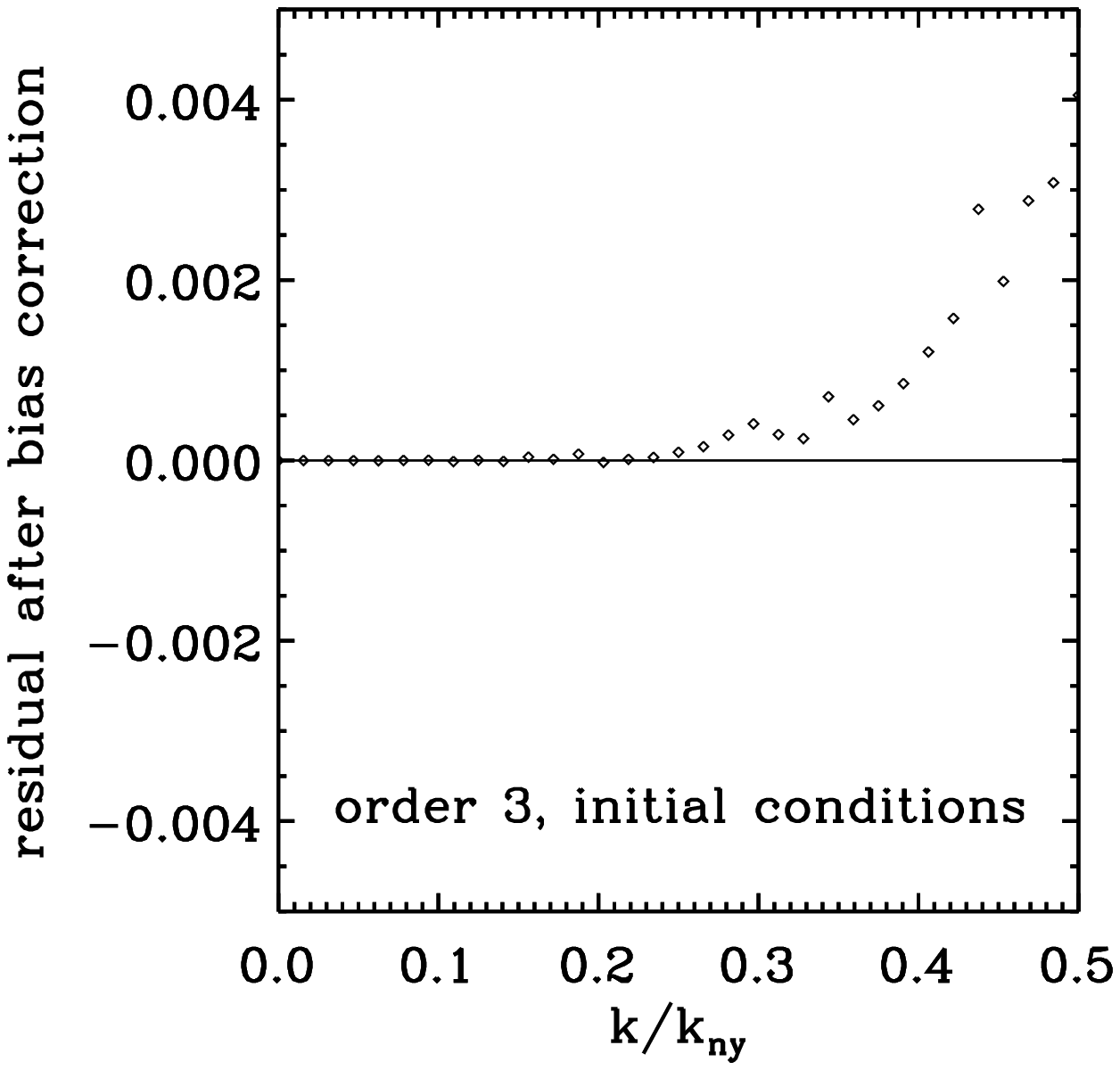,width=8cm}
}}
\caption[]{A zoom in the region $[0,k_{\rm ny}/2]$ of the relative residual,
$\delta P(k)/P(k)=[P_{\rm est}^{(N)}-P(k)]/P(k)$, 
on the unbiased third-order power spectrum, as measured
in our CDM {\tt GADGET} sample. The upper panel corresponds to present time,
where shot-noise correction was performed,
while the lower one corresponds to initial conditions, with
no shot noise correction. }
\label{fig:zoom0v5ny}
\end{figure}
In the next section, we shall  present a procedure to extend the
dynamic range of available values of $k$, which has been restricted
up to now the the Nyquist range of the (arbitrary) sampling grid used to perform
the Fourier-Taylor transform. This method will allow us  to 
control the accuracy of the measurements at all wavenumbers,
except of course those corresponding to modes close to the box size.
The previous analyses suggest keeping $k$ far away
enough from $k_{\rm ny}$. The choice proposed in this paper
is $k \leq k_{\rm ny}/2$. Figure \ref{fig:zoom0v5ny} shows the
relative residual on the third-order unbiased estimator in the region $k \in [0,k_{\rm
    ny}/2]$. Such a residual is well below the statistical errors,
of the order of $1.5\times 10^{-4}$ and $4\times 10^{-3}$ at most
in the final and the initial stage of the simulation respectively.
This error, already small, can be reduced further by increasing
the order $N$. However, it does not make sense to do so as long as
the statistical errors are dominant: to reduce the statistical errors
one has to sample more modes, \emph{i.e.}, to increase the size of the grid
used to perform the Fourier-Taylor transform. This is discussed
in next section. 
\section{Arbitrary wavenumbers with controlled error}
\label{sec:extension}
In this section, we show how the dynamical range of available values
of $k$ can be increased arbitrarily while keeping the statistical
errors and the aliasing effects on the measurements bounded 
(Jenkins et al., 1998).
 
As shown above, the error on the Fourier-Taylor expansion is basically 
controlled by the magnitude of $k$: this latter has to remain a small enough fraction
of the Nyquist frequency to avoid uncontrollable effects of aliasing
and hence to be able to maintain sufficiently low order $N$. 
Assuming that $\delta_k$ was computed in a given range of values of
(each coordinate of) $k$, $[-\alpha k_{\rm ny},\alpha k_{\rm ny}]$,
$\alpha \leq 1$, one can now consider an arbitrary shift,
$k_s$, such that $k+k_s$ is outside the available range,
\emph{e.g.}, $k_s=2 \alpha k_{\rm ny}$ to have access to the range $[\alpha k_{\rm ny},
  3\alpha k_{\rm ny}]$ in 1D.
Equation (\ref{eq:distra}) is rewritten 
\begin{eqnarray}
\delta_{k+k_s} & = & \frac{1}{N_{\rm p}} \sum_i M_i \exp(I k\cdot x_i)
\exp(I k_s \cdot x_i), \\
 & = & \frac{1}{N_{\rm p}} \sum_i M'_i \exp( I k\cdot x_i ), \label{eq:foldinga1}
\end{eqnarray}
with
\begin{equation}
M'_i=M_i \exp(I k_s \cdot x_i).
\label{eq:foldinga2}
\end{equation}
One thus obtains, by simple multiplication of the weights
by $\exp(I k_s.x_i)$, a modification of the Fourier-Taylor algorithm
that gives access to arbitrary values of $k$ while maintaining the
errors bounded. Of course, a new set of $N_{\rm FFT}$ transforms has
to be performed for each value of $k_s$.  

While this method allows us to investigate arbitrary values of $k$,
there is still the limitation imposed  by the available computer resources.
For instance increasing the computing volume from a cube
of sides $[-k_{\rm ny},k_{\rm ny}]$ to a cube of sides
$[-M\ k_{\rm ny}, M\ k_{\rm ny}]$ amounts to a calculation $M^D$ times
more expensive than for the original data cube. 
Instead, one can notice the following property of Fourier transform 
in one dimension. Writing
\begin{eqnarray}
\delta_{2k}& = & \frac{1}{N_{\rm p}} \sum_i \exp(2 Ik x_i) \\
 &=&\frac{1}{N_{\rm p}}
\sum_{i,x_i \in [0,L/2[} \exp( 2 I k x_i ) + \nonumber \\
& & +  \frac{1}{N_{\rm p}}
\sum_{i,x_i \in [L/2,L[} \exp[2 I k (x_i-L/2) + I k L ],
\end{eqnarray}
we can set 
\begin{eqnarray}
r_i & = & 2 x_i\ {\rm for}\ x_i \in [0,L/2[, \nonumber\\
r_i & = & 2 x_i-L\ {\rm for}\ x_i \in [L/2,L[. \label{eq:foldpart}
\end{eqnarray}
 We then have
\begin{equation}
\delta_{2k}= \frac{1}{N_{\rm p}} \sum_i \exp( I k r_i ),
\label{eq:folding}
\end{equation}
if we assume that $k L/(2\pi)$ is an integer; in that case
$\exp( I k L)=1$. 
Similarly one can write
\begin{equation}
\delta_{2k+1}=\frac{1}{N_{\rm p}} \sum_i S(i) \exp( I k r_i), 
\label{eq:folding2}
\end{equation}
where the sign function $S(i)$ satisfies
\begin{eqnarray} 
S(i) & = &\,\,\, \,\,1, \,\, {\rm if} \,\, x_i \in [0,L/2[,\\
S(i) & = & -1, \,\, {\rm if} \,\, x_i \in [L/2,L[.
\end{eqnarray}
This means that we  now have access to a doubly-large range of integer values
of $k L/(2\pi)$ by applying Fourier-Taylor method on simple foldings of the particle
distribution. Note that periodicity is not assumed here, except
that only the  \emph{integer} values of $k L/(2\pi)$ are available.
Such a set is complete in the periodic case.

\begin{table}
\caption[]{The sampled values of $k L/(2\pi)$ when using
the folding procedure explained in the main text on a grid
with $N_{\rm g}=16$. The trivial fundamental case, $k=0$,
is not shown here. From the left column to the right one,
one considers $M=0$ folding to $M=4$ foldings. Each column
defines a range of available values of $k$. 
As discussed in the text, there can be several estimates of $P(k)$ for a given value
of $k$. For instance we have 3 estimates of  $P(4)$.
To make unknown aliasing effects always negligible in practice
compared to statistical errors, we impose $k \leq k_{\rm
  ny}^{(M)}/2$, where $k_{\rm  ny}^{(M)}$ is the effective wavenumber 
probed by the Nyquist frequency of the grid after $M$ foldings (this is indicated
by the mid-horizontal line on the table). To minimize the statistical
errors, $M$ should then be as small as possible, since this parameter
controls the sparseness of sampling in Fourier space. 
To measure $P(4)$ for instance, one takes
$M=0$. In detail, the sampled values of $k$
are highlighted in
bold on the table, $k=1,2,4,6,8,12,16,24,32,48,64$;
Fourier space is increasingly sparse sampled with $k$.
This sparse sampling roughly increases linearly in logarithmic
scale, but the width of the bin used to measure $P(k)$ increases
likewise,  keeping the number of sampled independent modes $C(k)$ bounded
between two values $C_1$ and $C_2$ independent of the number of foldings $M$,
except for $M=0$ [here $C_1=98$ and $C_2=210$ for $D=3$].
As a result, the statistical errors on the power spectrum measurement
remain bounded as well, as illustrated by Fig.~\ref{fig:errors}.
}
\vskip 0.5cm
\begin{tabular}{rrrrr|c}
$M=0$ & $M=1$ & $M=2$ & $M=3$ & $M=4$ & $C(k)\ (D=3)$ \\
\hline
{\bf 1}  &  2        &  4        &  8         &  16 &  18 \\
{\bf 2}  &  4        &  8        &  16        &  32 &  62 \\
{\bf 3}  &  {\bf 6}  & {\bf 12}  & {\bf 24}   &  {\bf 48} &  98 \\
{\bf 4}  &  {\bf 8}  & {\bf 16}  & {\bf 32}   &  {\bf 64} & 210 \\
\hline
5  & 10  & 20  & 40   &  80 & 350 \\
6  & 12  & 24  & 48   &  96 & 450 \\
7  & 14  & 28  & 56   & 112 & 602 \\
8  & 16  & 32  & 64   & 128 & 762 \\
\hline
\end{tabular}
\label{tab:tablefold}
\end{table}
Obviously this folding trick can be generalized to higher number of
dimensions. In that case, the number of Fourier transforms needed to
perform the calculations increases by
a factor $2^D$ each time a factor 2 is gained in the dynamic range
of available values of $k$, exactly as in 
equations~(\ref{eq:foldinga1}) and (\ref{eq:foldinga2}),
as expected. However, we can restrict here to the simplest
folding given by equation~(\ref{eq:folding}) and ignore foldings involving equation~(\ref{eq:folding2}). 
This means that we are {\em
  sparse-sampling} Fourier space, increasing 
the dynamic range  by a factor two each time, maintaining the
computational time of the same order at each step of the
procedure.\footnote{Some of the possible caveats of such a sparse
sampling are discussed in Jenkins et al.~(1998).}
Table~\ref{tab:tablefold} shows the list of values
of $k$ sampled in one dimension in a simple case, where $N_{\rm
  g}=16$. Note that the  number of available modes per sampled
value of $k$ remains the same as for the original sampling: the
error on the rough power spectrum shall remain bounded as
we discuss later below.

In practice the folding algorithm works as follows:
\begin{enumerate}
\item[(i)] First apply the Fourier-Taylor estimator on the original particle
distribution sampled on the periodic grid of size $N_{\rm g}$, 
and measure $P_{\rm est}^{(N)}(k)$ up to some fraction of
the Nyquist frequency, $\alpha k_{\rm ny}$, with
\emph{e.g.}, $\alpha=1/2$ as advocated in the previous
section. That corresponds to the first step of the algorithm, with
zero folding, $M=0$.
\item[(ii)] (Assuming we are at step $M$ of the process.) Fold the
particles in each dimension according to equation~(\ref{eq:foldpart}), 
resample them again on a periodic
grid of size $N_{\rm g}$, but which probes a physical size twice
smaller than in the previous step, $L_{M+1}=L_{M}/2=L_0/2^{M+1}$ ($L_0=L$). 
Measure $P_{\rm est}^{(N)}(k)$
up to $\alpha$ times the Nyquist frequency of that grid, which in practice
corresponds to twice the Nyquist frequency of the grid of the previous
step, $k_{\rm ny}^{(M+1)}=2k_{\rm ny}^{(M)}=2^{M+1} k_{\rm ny}$. 
\item[(iii)] Repeat step (ii),
until the value $\alpha k_{\rm ny}^{(M+1)}$ is as large as
required, for instance until the softening scale of the simulation has
been reached.
\end{enumerate}
We now have a set of values of measurements of the power spectrum
for a number of ranges of values of $k$ as for instance illustrated by
Table~\ref{tab:tablefold}. These ranges overlap from one folding to
another: this can be used to check for systematic errors on the
measurements. Indeed, for one value of $k$, there can be several
measurements of $P(k)$. One has, for each folding, to choose
the range of values of $k$ that contribute to the final measurement.
To do that, one could think of compromising
between the statistical errors, which are larger when $k$ is small,
 and the unknown systematics brought by aliasing, which can
become significant when $k$ approaches the Nyquist frequency of
the grid. The choice of compromise depends on the order 
$N$ of the Fourier-Taylor transform considered. It is theoretically possible to
 find an ``optimal'' compromise between the parameter $\alpha$, the
order $N$ and the resolution of the grid, $N_{\rm g}$, to 
maintain the errors on the measurement
of $P(k)$ below a given limit at a minimum computational cost. 
However, such a project would go beyond the scope of this paper.
Our strategy here is rather to make sure that the
unknown systematics due to aliasing are negligible compared to the
statistical error given by \emph{e.g.}, equation~(\ref{eq:errgaus}). 
We therefore advocate $\alpha=1/2$ and sufficiently
high order Fourier-Taylor approximation. For instance, the third-order transform
should in practice do well enough for $N_{\rm g}$ of the order
or smaller than a thousand.

\begin{figure*}
\centerline{\hbox{
\psfig{file=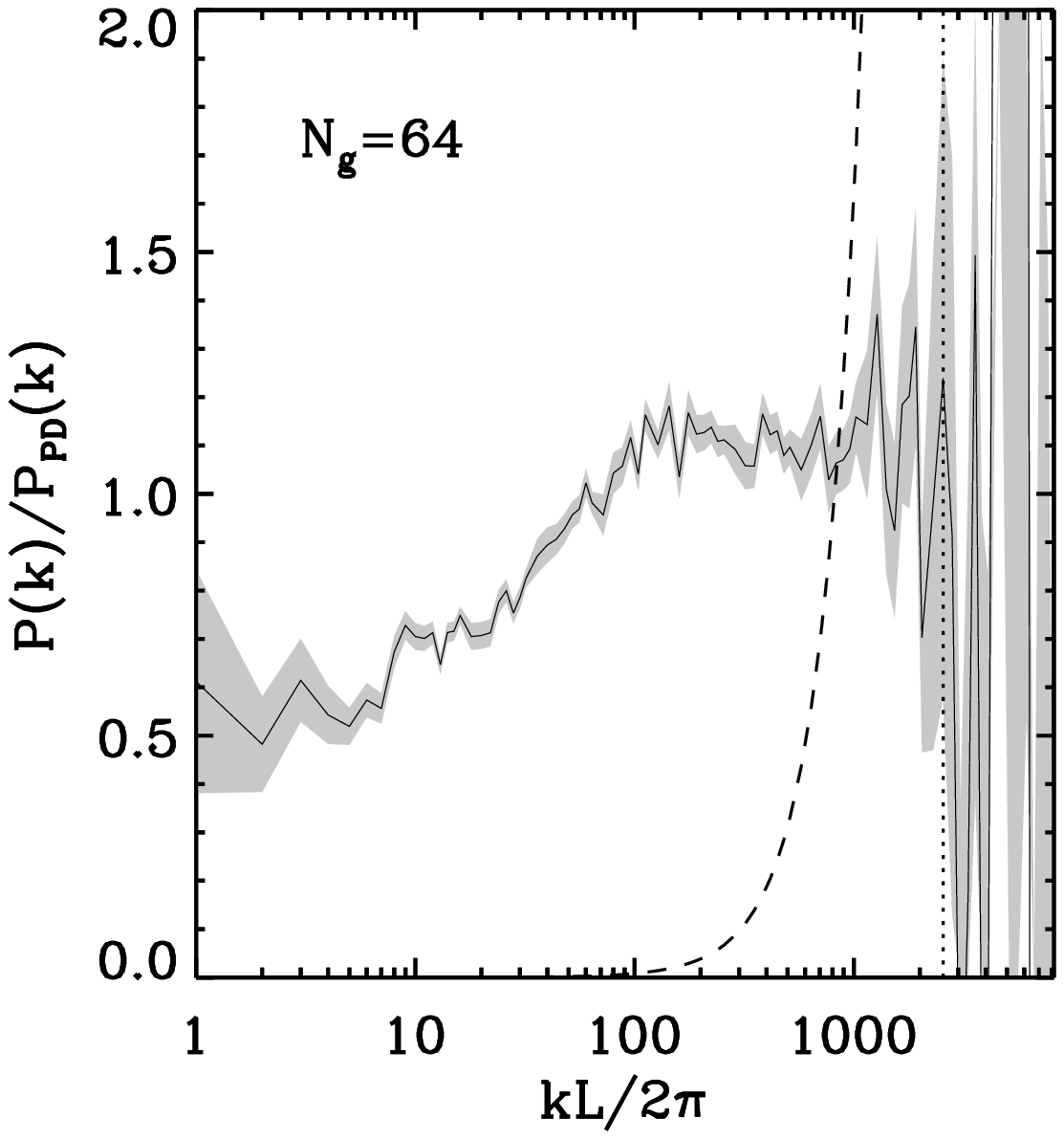,width=8cm}
\psfig{file=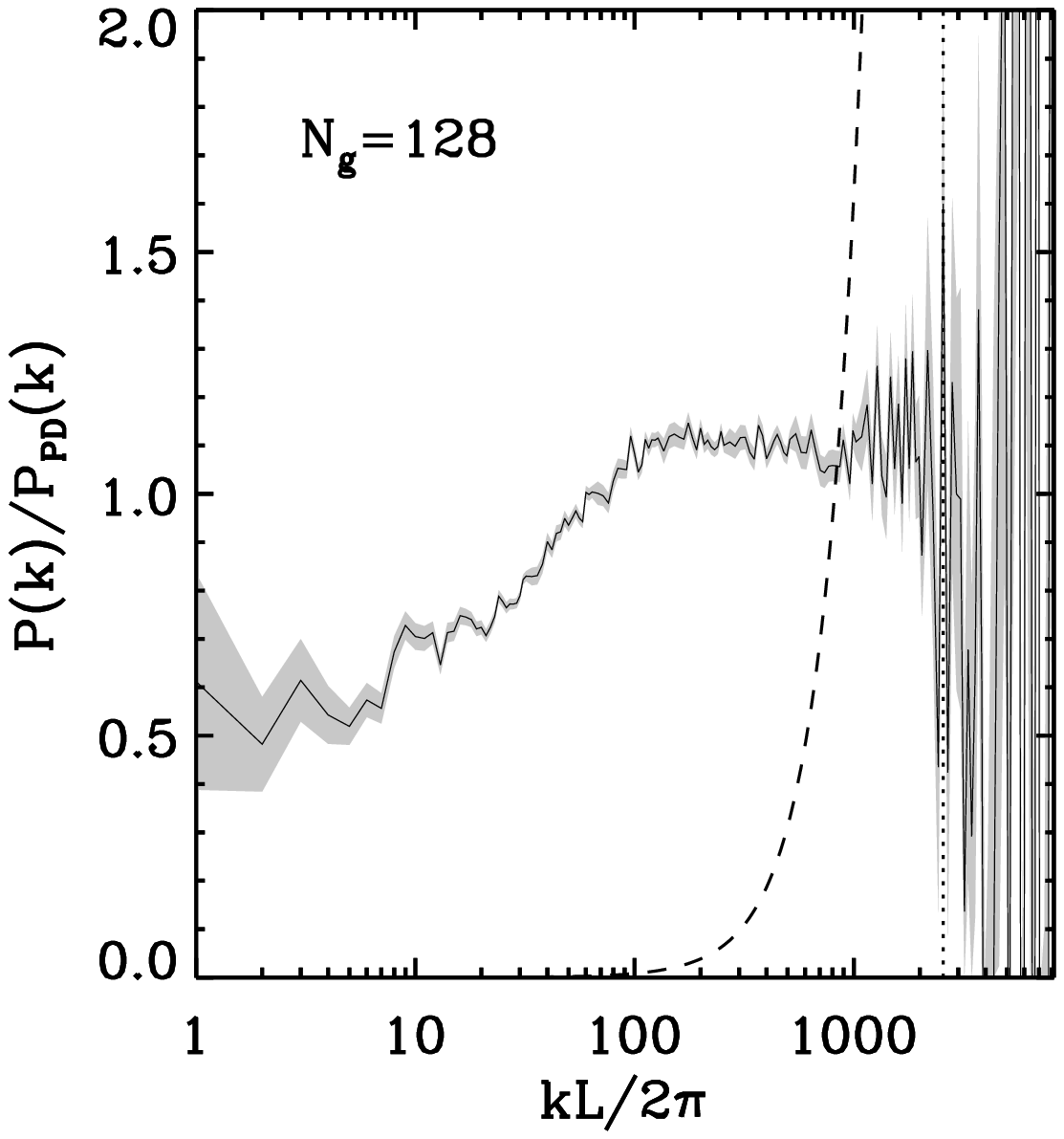,width=8cm}
}}
\centerline{\hbox{
\psfig{file=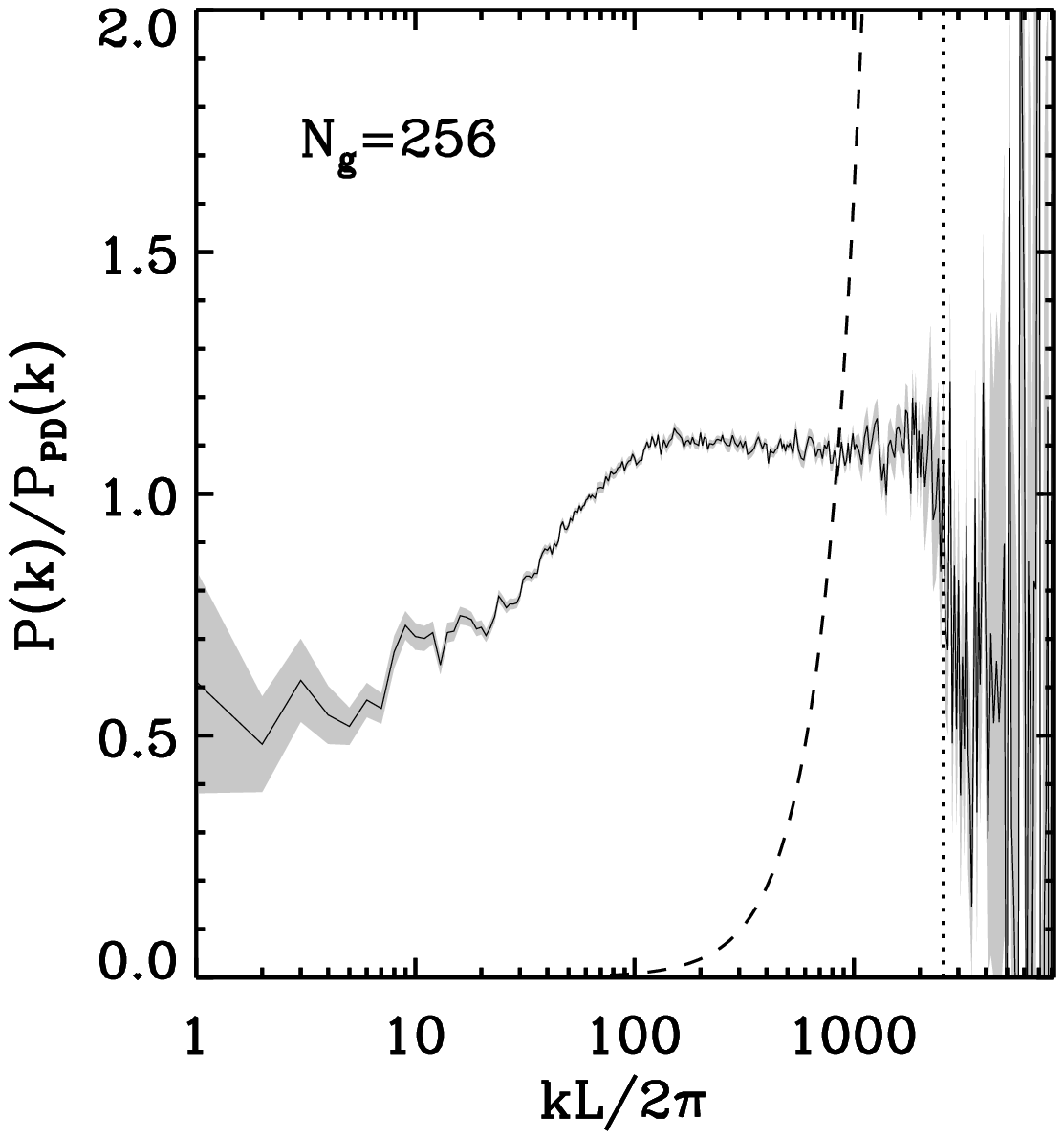,width=8cm}
\psfig{file=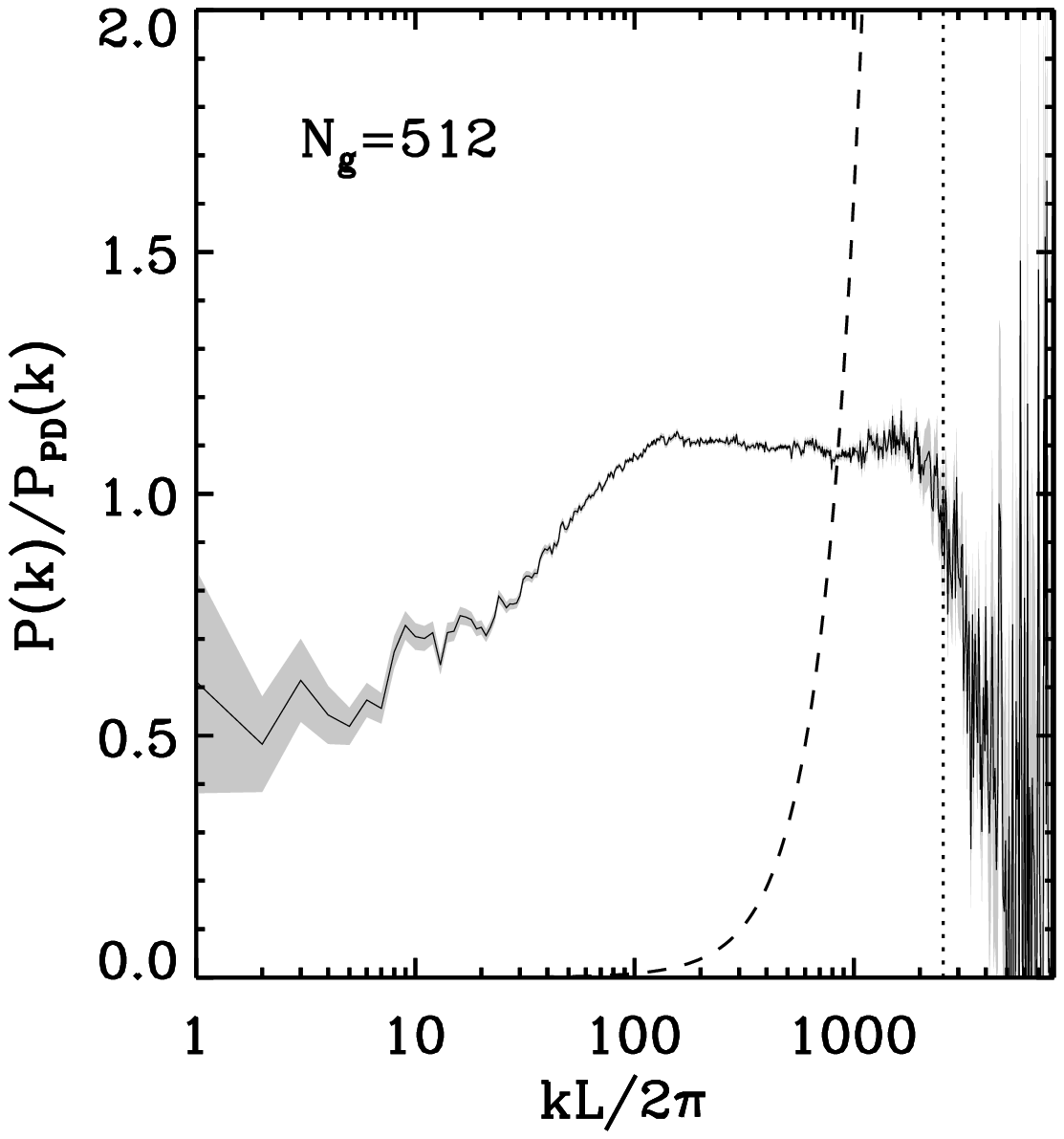,width=8cm}
}}
\caption[]{The power spectrum measured at present time 
in our CDM {\tt GADGET} sample, with
the estimator $P^{(3)}_{\rm est}(k)$ (equation~\ref{eq:estimator2}), using
the folding algorithm explained in the main text to (sparsely) probe  the full dynamic range $k=[1,8192]$. To see the details better,
the measurements have been divided by the analytical proxy $P_{\rm
  PD}(k)$ of Peacock \& Dodds (1996). Each panel corresponds to
a choice of the grid resolution, $N_{\rm g}$, used to perform the
Fourier-Taylor algorithm in combination with the folding of the
particle distribution. 
The measurements are represented
by the thin curve, while the shaded region gives the statistical
error estimated with equation~(\ref{eq:errorint}). The dashed curve
corresponds to the shot noise of the particles. The dotted vertical line indicates the softening
length $\varepsilon$: above that scale, $P(k)$ should present a cut-off,
which is clearly visible when the signal-to-noise ratio is large
enough, see \emph{e.g.}, the lower-right panel.
}
\label{fig:reseffect}
\end{figure*}
\begin{figure*}
\centerline{\hbox{
\psfig{file=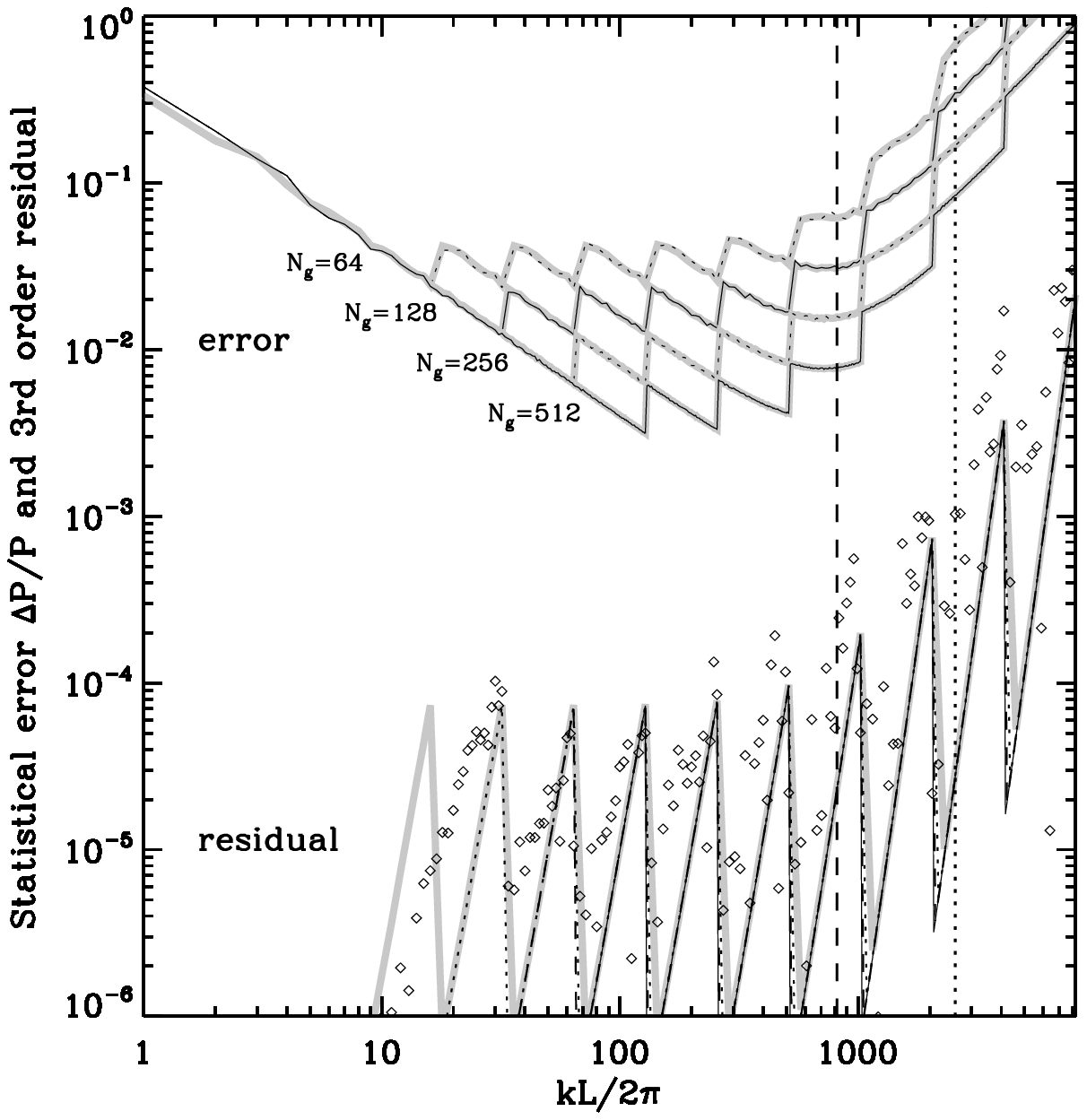,width=14cm}
}}
\caption[]{The statistical error on the measurement of $P(k)$ when
using the folding procedure explained in the main text. The upper
curves correspond to the estimated statistical relative error on $P^{(3)}_{\rm
  est}(k)$ (but these curves would not change significantly for other
values of the order $N$ of the Fourier-Taylor expansion): 
for various values of $N_{\rm g}$ as indicated
on the plot, the thin (alternatively dotted and solid) 
and the thick grey curves  correspond respectively
to eqs.~(\ref{eq:errorint}) and (\ref{eq:errgaus}) for estimating the errors on the
measurement of $P(k)+1/N_{\rm p}$.
The function represented here is given in fact by
equation~(\ref{eq:erroreff}) as discussed in detail in the main text.
For comparison, the lower curves give an estimate of the expected
residual due to aliasing effects on the power spectrum, while the
symbol show the measurement of these residuals for $N_{\rm g}=128$.
}
\label{fig:errors}
\end{figure*}

Figure~\ref{fig:reseffect} illustrates how the measurements behave over
the range $k L/(2\pi) \in [1,8192]$ when one changes the
grid resolution from $N_{\rm g}=64$ to $N_{\rm g}=512$ in the folding
algorithm.
In order to see the details better, the function represented on each
panel is $P^{(3)}_{\rm est}(k)/P_{\rm PD}(k)$,
where $P_{\rm PD}(k)$ is the function given by the nonlinear ansatz of
Peacock \& Dodds (1996). This figure is supplemented with
Fig.~\ref{fig:errors}, which gives for each case
considered in Fig.~\ref{fig:reseffect} the errors on $P^{(3)}_{\rm
  est}(k)$, in the following way to make the plot readable: the
thin curve (alternatively dotted and continuous) and the thick grey curves on the upper
part of Fig.~\ref{fig:errors} correspond to equations~(\ref{eq:errorint}) and
(\ref{eq:errgaus}) respectively. However these equations give the error on
the estimate of the power spectrum plus the shot noise contribution
of the particles. In order to get an estimate of the statistical error
on $P(k)$, we compute an expression similar 
to equation~(\ref{eq:errPoisson}), but as follows:
\begin{equation}
\frac{\Delta P}{P}  = \left\{ E(k)\ {\rm or} \ E_{\rm G}(k) \right\}
\times \frac{ P_{\rm PD}(k)+1/N_{\rm p} }{ P_{\rm PD}(k)}.
\label{eq:erroreff}
\end{equation}
That way, the plot is readable because we replace a noisy function with its
smooth guess $P_{\rm PD}(k)$.\footnote{If the
  error on the rough power spectrum is given by 
\begin{equation}
\left( \frac{{\Delta P}_{\rm rough}}{P_{\rm rough}} \right)^2=E^2,
\end{equation}
then it follows that the error on the shot-noise-corrected power
spectrum reads as in equation~(\ref{eq:errPoisson}) but with the term
$1/C(k)$ replaced with $E^2$. To obtain a smooth estimate of the errors
plotted on Fig.~\ref{fig:errors}, our choice is to replace $P(k)$ in 
equation~(\ref{eq:errPoisson}) with
its theoretical proxy, $P_{\rm PD}(k)$. After simple algebraic 
calculations, one just obtains equation (\ref{eq:erroreff}). 
In that framework, the expressions for the theoretical and
measured residuals on the power-spectrum estimator 
can also be written naturally as explained in the main text and
plotted on Fig.~\ref{fig:errors}.} 
The lower part of the same figure shows
the residual function due to aliasing, $R_3(k) \{ 1+1/[P_{\rm
  PD}(k) N_{\rm p}] \}$ (thick grey, dotted, dashed and solid lines, which
correspond to $N_{\rm g}=64$, 128, 256 and 512, respectively), 
as well as its measurement for $N_{\rm  g}=128$ (symbols). 
In the last case, the function displayed on
Fig.~\ref{fig:errors} is $[P^{(3)}_{\rm est}(k)-P(k)]/P_{\rm PD}(k)$.
It should roughly follow  the dotted line, which is indeed the
case. Note that a residual $f(k)=R_3(k)\{ 1+1/[P_{\rm
  PD}(k) N_{\rm p}] \}$ assumes $P_{\rm max}=P_{\rm PD}(k)$. 
Since $P_{\rm PD}(k)$ is a decreasing function of $k$, the function
$f(k)$ is expected to overestimate the true residual.\footnote{We
  assume here the framework of the assumptions of
  \S~\ref{sec:ftps}. For a perturbed grid, as studied in
\S~\ref{sec:badbad}, the residual are expected to be much larger,
as illustrated by the lower panel of Fig.~\ref{fig:zoom0v5ny}.} On the contrary,
the symbols tend to lie above the ``theory'', but this disagreement is
clearly within statistical errors, which is what really matters for the
point made here.

The mid-range values of $k$ in Fig.~\ref{fig:errors} show 
that increasing the resolution $N_{\rm g}$ of the grid by a factor of two improves the overall signal-to-noise ratio by a factor $2^{(D-1)/2}$, because $C(k)$ scales roughly like
$k^{D-1}$. This is illustrated as well by Fig.~\ref{fig:reseffect}, where the
small fluctuations of the measured spectrum at intermediate values of
$k$ decrease when $N_{\rm g}$ increases.  At small $k$, however, 
the errors are independent of $N_{\rm g}$ as
there are fewer and fewer statistically independent modes when one approaches the size
of the box, whatever the resolution of the grid used to perform the
measurements. At large $k$, one is dominated by the shot noise of the
particles: when $P(k) \la 1/N_{\rm p}$, as indicated by the dashed
line on Figs.~\ref{fig:reseffect} and \ref{fig:errors}, the error
on $P(k)$ increases dramatically, to become arbitrarily large when $P(k)$ is
subject to the hard cut-off due to the softening of the forces
at scales smaller than $\varepsilon$, as 
indicated by the dotted line on the figures. Indeed, in
the folding method proposed here, one is actually able to control the error 
on the measurement of the quantity $P(k)+1/N_{\rm p}$, and not on the measurement
of $P(k)$.  


\section{Summary}
\label{sec:conclusion}
In this paper we presented a method to estimate Fourier modes of a
particle distribution, based on a Taylor expansion of the
trigonometric functions. This idea is inspired from the work
of Anderson \&  Dahleh (1996). We paid particular attention to the
measurement of the power spectrum $P(k)$ when the point distribution
is the local Poisson realization of a stationary random field, where
explicit expressions for the ensemble average of a naive rough estimator
of $P(k)$ were derived. This allowed us to accurately determine  the
biases induced by discreteness and by the Taylor expansion, which
can be easily corrected for, and to
quantify the effects of aliasing, which are controlled by the order, $N$, of
the expansion. Our calculations show that effects of aliasing decrease
 quickly with $N$, as illustrated by Table~\ref{table:tableresi}.  
The analytic calculations were confronted successfully
with a cosmological $N$-body simulation. We also studied how the 
deviations from local Poisson behavior influence the measurements,
such as in initial conditions of the simulation, which correspond to
a perturbed grid pattern. We proposed an unbiased estimator which is
nearly free of aliasing, and which still performs well for the
perturbed grid. The accuracy of this estimator is thus entirely
controlled by the statistical errors, which arise from the finite
number of sampled modes. We also showed how the dynamical range in Fourier
space could be arbitrarily increased while keeping the statistical
error bounded, by appropriate foldings of the particle distribution,
as suggested by Jenkins et al.~(1998). 
Note that, while the Fourier-Taylor method was applied here to the
power spectrum, it can be easily generalized to higher order estimators, 
for instance to measure the bispectrum or the trispectrum of the
distribution. 

The Fourier-Taylor module as well as the associated power spectrum 
estimator tool we propose is available as an F90 package, {\tt powmes}, at {\tt
  www.projet-horizon.fr} or on request from the authors. 
It works with the {\tt GADGET} file format.

\section*{Acknowledgements}
We thank S. Prunet, K. Benabed and ``Colonel" R. Teyssier for useful discussions.
This work was completed in part as a task of the HORIZON project
({\tt www.projet-horizon.fr}) and was supported by STFC in the UK. 

\appendix
\section{Ensemble averages with various assumptions}
\label{sec:appendix}
There are subtleties that arise when it comes to performing ensemble
averages. While this has been widely discussed in the literature (see,
\emph{e.g.}, Peebles 1980), we address the issue briefly again here.
Given a distribution of $N_{\rm p}$ particles in a
(hyper)cubical volume $V$, which is a discrete realization
of an underlying random continuous field, $\rho(x)$ (of average unity),
one measures the two quantities 
\begin{equation}
F=\sum_{i=1}^{N_{\rm p}} f(x_i),\quad G=\sum_{i=1}^{N_{\rm p}} g(x_i),
\label{eq:defFG}
\end{equation}
where $f$ and $g$ are some functions. 
The question
we want to address here is how to compute the ensemble average of $FG$ over many
realizations of the underlying distribution, given some constraints. 
There are two relevant cases to consider:
\begin{enumerate}
\item {\em The ``realistic'' case:}  $V$ is a subvolume of a realization of
much larger volume. In that case ensemble average allows the number of
particles $N_{\rm p}$ to vary, as well as the average density over the
volume: the quantity
\begin{equation}
\rho(V)=\frac{1}{V} \int_V \rho(x) d^Dx,
\label{eq:rhodev}
\end{equation}
is allowed to fluctuate around the mean.
\item {\em The $N$-body simulation standard case:} in that case,
  $N_{\rm p}$ is fixed, as well as $\rho(V) \equiv 1$. 
\end{enumerate}
\subsection{The ``realistic'' case: unconstrained ensemble average}
\label{sec:a1}
Following Peebles (1980), we divide $V$ into infinitesimal cells of
volume $\delta V$, such that they contain zero or one particle.
Let $n$ be the number of particles they contain, and ${\bar n}=\langle
n \rangle$ its ensemble average. Let $p(n,x)$ be the probability of
having $n$ particles in an infinitesimal cell at position $x$. Then, the local random
process characterizing the realization of the smooth field in a
distribution of particles gives, assuming that $\langle \rho \rangle=1$,
\begin{eqnarray}
p(1,x) &= & {\bar n} \delta V \rho(x), \\
p(0,x) &=  & 1-{\bar n} \delta V \rho(x).
\end{eqnarray}
The sum (\ref{eq:defFG}) can be rewritten over the infinitesimal cells,
labelled as $j$ and at positions $c_j$,
\begin{equation}
F=\sum_j n_j f(c_j), \quad G=\sum_j n_j g(c_j).
\end{equation}
So
\begin{equation}
\langle F \rangle=\sum_j \langle n_j f(c_j) \rangle=\sum_j {\bar n}
\delta V \langle \rho \rangle f(c_j),
\end{equation}
which gives in integral notation
\begin{equation}
\langle F \rangle={\bar n} \int_V d^Dx f(x),
\label{eq:FAV}
\end{equation}
and likewise for $G$.
The product, $FG$, is then
\begin{equation}
FG=\sum_j n_j^2 f(c_j)g(c_j) + \sum_{j\neq j'} n_j n_{j'} f(c_j)g(c_j').
\end{equation}
\begin{eqnarray}
\langle FG \rangle & = & {\bar n} \int_V f(x)g(x)d^Dx  + \nonumber \\
 & & + \ {\bar n}^2
\int_{V} [1+\xi(x,y)]  f(x)g(y) d^Dxd^Dy.
\label{eq:FGuncon}
\end{eqnarray}
In this equation,  we have defined the two-point
correlation  function
\begin{equation}
\langle \rho(x)\rho(y) \rangle \equiv 1 + \xi(x,y).
\label{eq:xi2df}
\end{equation}
In this paper, we assume stationarity: $\xi(x,y)=\xi(x-y)$.
Equation (\ref{eq:FGuncon}) is the basis that we used for the
calculation of the ensemble average of $P^{(N)}(k)$ in \S~\ref{sec:powspecesti},
and from which we derive equation~(\ref{eq:finalres}).
\subsection{The $N$-body simulation case: constrained ensemble
  average }
\label{sec:a2}
Since $N_{\rm p}$ is now fixed, the method of infinitesimal cells does
not work anymore, at least not straightforwardly. 
However, what still remains valid is that the probability
of having a particle at position $x$ is proportional to $\rho(x)$.
More generally, the  probability density of having a set of particles 
at positions $(x_1,\cdots,x_{N_{\rm p}})$, given the realization
$\rho(x)$, is given by
\begin{equation}
p(x_1,\cdots,x_{N_{\rm p}})=\frac{1}{V^n}
\rho(x_1)\cdots\rho(x_{N_{\rm p}}),
\end{equation}
remembering that $\rho(V)$ (equation~\ref{eq:rhodev}) is now constrained to
be unity for each realization of the ensemble average.
Then
\begin{eqnarray}
\langle F \rangle & = & \sum_{i=1}^{N_{\rm p}} \left\langle \int f(x_i)   p(x_1,\cdots,x_{N_{\rm
    p}}) d^D x_1\cdots d^D x_{N_{\rm p}} \right\rangle \nonumber \\  &
    = & \frac{N_{\rm p}}{V}
    \int_V f(x)d^Dx,
\end{eqnarray}
after performing the integrals and then ensemble averaging, 
using $\langle \rho \rangle=1$. 
As a result we converge again to equation~(\ref{eq:FAV}), since ${\bar
  n}=N_{\rm p}/V$.  
However, the calculation of $\langle FG \rangle$ gives, 
\begin{eqnarray}
\langle FG \rangle =  \frac{N_{\rm p}}{V} \int_V  f(x)g(x) d^Dx 
 +   \frac{N_{\rm p} (N_{\rm p}-1)}{V^2} \times \nonumber \\
 \times \int_V [1+\xi(x,y)]
f(x)g(y) d^Dx d^Dy,
\end{eqnarray}
using the definition (\ref{eq:xi2df}).
The second term of this expression differs from second term of equation~(\ref{eq:FGuncon}),
by a factor $(N_{\rm p}-1)/N_{\rm p}$. In addition, the constraint
$\rho(V)=1$ for each realization reads, after ensemble averaging,
\begin{equation}
\int_V \xi(x,y) d^Dx d^Dy =0.
\end{equation}
These differences impose, in particular, that the fundamental mode,
$P^{(N)}(0)$ is always exactly unity for an $N$-body simulation in our
choices of units, unlike equation~(\ref{eq:finalres}), which was computed using the
method explained in Appendix A1.
\section{Some useful analytic expressions}
The calculation of the number $\kappa_{n}(k,M)$ (equation~\ref{eq:kapan})
can be performed
easily by using the multinomial theorem:
\begin{eqnarray}
\kappa_{n}(k,M)=\sum_{q_1+\cdots+q_D=n} \frac{n!}{q_1!\times \cdots \times
  q_D!} \times \nonumber \\ \times \eta_{q_1}(k_1,M_1)\times \cdots \times \eta_{q_D}(k_D,M_D),
\end{eqnarray}
where
\begin{equation}
\eta_{n}(k,M)=\int_{-1/2}^{1/2} \exp[-I\ (k+2\pi M)\Delta]\ (I\ k \Delta)^n d\Delta
\label{eq:etan}
\end{equation}
is similar to $\kappa_{n}(k,M)$ but is computed on a scalar instead of a
vector: for $D=1$, $\eta_{n}(k,M)=\kappa_n(k,M)$.
Note that $\eta_{n}(k,M)$ can be computed using the following
recursion:
\begin{equation}
\eta_0(k,M)=(-1)^M \sin(k/2)/[(k+2\pi M)/2],
\end{equation}
\begin{eqnarray}
\eta_n(k,M)=\frac{(-1)^M}{k+2\pi M} ( {k}/{2} )^{n} I \{ I^n
  \exp( -I\ k/2)- \nonumber \\ -\ I^{-n} \exp( I\ k/2) \} +
  \frac{n\ k}{k+2\pi M} \eta_{n-1}(k,M).
\end{eqnarray}
The final result can be expressed as
\begin{eqnarray}
\eta_n(k,M)  =  \frac{2(-1)^M k^n n!}{(k+2\pi M)^{n+1}} \times \nonumber
\\  \times\ \left\{
\sin(k/2) \sum_{l=0}^{n/2} \frac{(-1)^l}{(2l)!}(k/2+\pi M)^{2l}  - \right.
\nonumber \\
 \left.-\ \cos(k/2)  \sum_{l=0}^{(n-1)/2} \frac{(-1)^{l}}{(2l+1)!} (k/2+\pi
M)^{2l+1}  \right\}.
\label{eq:etan2}
\end{eqnarray}

\section{The power spectrum of a perturbed grid}
\label{sec:appendixB}
We consider here the case of a three-dimensional grid pattern of particles perturbed
by a Gaussian random displacement, which is curl-free, stationary and
isotropic.

Isotropy and stationarity imply that the joint probability distribution of displacements
${\cal P}_1={\cal P}(q_1)$, ${\cal P}_2={\cal P}(q_2)$ depends only on $q_{12}=|q_1-q_2|$ and gives
\begin{eqnarray}
{\cal L}({\cal P}_1,{\cal P}_2,q_{12}) & = & \frac{27}{(2\pi)^3 (1-\rho^2)^{3/2}
  \sigma^6}\times \nonumber \\ & & 
 \times\ \exp\left[ - \frac{{\cal P}_1^2+{\cal P}_2^2-2 \rho {\cal P}_1 \cdot {\cal P}_2}{2 \sigma^2
    (1-\rho^2)/3} \right],
\end{eqnarray}
where
\begin{equation}
\sigma^2=\langle {\cal P}_1^2 \rangle=\langle {\cal P}_2^2 \rangle, 
\quad \rho(q_{12}) \sigma^2= \langle {\cal P}_1 \cdot {\cal P}_2 \rangle,
\end{equation}
are the variance of the displacement field and its correlation
function, respectively.  The Fourier modes of the perturbed grid
pattern are given by equation~(\ref{eq:Fouriergrid}).
The constrained ensemble average  of the power spectrum estimate 
(keeping $N_{\rm p}$ fixed; see Appendix A2) is
\begin{eqnarray}
\langle \delta_k \delta_{-k} \rangle  =  \delta_{\rm
  D}(0)+\frac{1}{N_{\rm p}}+ \frac{1}{N_{\rm p}^2} 
  \sum_{q_1 \neq q_2}\exp[ I k \cdot (q_1-q_2)] \times \nonumber \\ \times  
\langle \exp\{ I k\cdot [{\cal P}(q_1)-{\cal P}(q_2)]\} \rangle. \\
  =  \delta_{\rm D}(0)+\frac{1}{N_{\rm p}} +\frac{1}{N_{\rm p}^2}
  \sum_{q_1 \neq q_2} \exp[ I k  \cdot (q_1-q_2)] \times \nonumber \\
  \times \int d^3{\cal P}_1 d^3{\cal P}_2 {\cal L}({\cal P}_1,{\cal P}_2,q_{12}) \exp[ I k  \cdot({\cal P}_1-{\cal P}_2)
  ].
\end{eqnarray}
Notice that the offset $s$ has disappeared from this expression, as expected.
After some algebra, one finds equation~(\ref{eq:powergrid}),
where the diagonal term has been trivially integrated with the
off-diagonal one. This expression gives the discrete version of the Zel'dovich
power spectrum (see Schneider \& Bartelmann, 1995, for the continuous limit).



\begin{thebibliography}{}
\bibitem{} Abazajian K., Zheng, Z., Zehavi, I., Weinberg, D. H.,
  Frieman, J. A., Berlind, A., Blanton, M. R., Bahcall, N. A.,
  Brinkmann, J., Schneider, D. P., Tegmark, M., 2005, ApJ 625, 613
\bibitem{} Anderson, C., Dahleh, M., 1996, SIAM J. Sci. Comput.
  17, 913
\bibitem{} Bardeen, J. M., Bond, J. R., Kaiser, N., Szalay, A. S.,
  1986, ApJ 304, 15
\bibitem{} Baumgart, D. J., Fry, J. N., 1991, ApJ 375, 25
\bibitem{} Benjamin, J., et al., 2007, MNRAS 381, 702 
\bibitem{} Bernardeau, F., Colombi, S., Gazta\~naga, E., Scoccimarro,
  R., 2002, PhR 367, 1
\bibitem{} Bertschinger, E., 2001, ApJS 137, 1
\bibitem{} Cooray, A., Sheth, R., 2002, PhR 372, 1
\bibitem{} Crocce, M., Pueblas, S., Scoccimarro, R., 2006, MNRAS 373, 369
\bibitem{} Croft, A. C., Weinberg, D. H., Pettini, M., Hernquist,
  L., Katz, N., 1999, ApJ 520, 1 
\bibitem{} Cui, W., Liu, L., Yang, X., Wang, Y., Feng, L., Springel,
  V., 2008, submitted to ApJ (arXiv:0804.0070)
\bibitem{} Daubechies, I., 1988, Comm. Pure Appl. Math., 41 (7), 909
\bibitem{} Dunkley, J., et al., 2008, ApJS, in press
  (arXiv0803.0586)
\bibitem{} Feldman, H. A., Kaiser, N., Peacock, J. A., 1994, ApJ 426,
  23
\bibitem{} Fu, L., et al., 2008, A\&A 479, 9
\bibitem{} Gabrielli, A., 2004, PhRvE 70, 066131
\bibitem{} Hamilton, A. J. S., Kumar, P., Lu, E., Matthews A., 1991,
  ApJ 374, L1
\bibitem{} Hamilton, A. J. S., Rimes, C. D., Scoccimarro, R., 2006,
  MNRAS 371, 1188
\bibitem{} Hockney, R. W., Eastwood, J. W., 1988, Computer Simulation
  Using Particles (Institute of Physics Publishing, Bristol and
  Philadelphia)
\bibitem{} Jenkins, A., Frenk, C. S., Pearce, F. R., Thomas, P. A., 
  Colberg, J. M., White, S. D. M., Couchman, H. M., Peacock, J. A.,
  Efstathiou, G., Nelson, A. H., 1998, ApJ 499, 20
\bibitem{} Jing, Y. P., 2005, ApJ 620, 559
\bibitem{} Joyce, M., Marcos, M., 2007a, PhRvD 75, 063516
\bibitem{} Joyce, M., Marcos, M., 2007b, PhRvD 76, 103505
\bibitem{} Ma, C.-P., Fry, J. N., 2000, ApJ 543, 503
\bibitem{} Marcos, B., Baertschinger, T., Joyce, M., Gabrielli, A.,
  Sylos Labini, F., 2006, PhRvD 73, 103507
\bibitem{} Martinez, V. J., 2008, to appear in Data Analysis in Cosmology, 
Lecture Notes in Physics, 2008, eds. V. J. Martinez, E. Saar,
E. Martinez-Gonzalez, \& M.J. Pons-Borderia, Springer-Verlag
(arXiv:0804.1536)
\bibitem{} Peacock, J. A., Dodds, S. J., 1996, MNRAS 280, L19
\bibitem{} Peacock, J. A., Smith, R. E., 2000, MNRAS 318, 1144
\bibitem{} Peebles, P. J. E., 1980, The Large Scale Structure of The
  Universe (Princeton University Press, 1980)
\bibitem{} Potts, D., Steidl G., Tasche, M., 2001, Fast Fourier
transforms for nonequispaced data: A tutorial, in 
Modern Sampling Theory: Mathematics and Applications,
eds J. J. Benedetto \& P. Ferreira, Chapter 12, p. 249
\bibitem{} Rimes, C. D., Hamilton, A. J. S., 2006, MNRAS 371, 1205
\bibitem{} Scoccimarro, R., Colombi, S., Fry, J. N., Frieman, J. A.,
  Hivon, E., Melott, A., 1998, ApJ 496, 586
\bibitem{} Scoccimarro, R., Sheth, R. K., Hui, L., Jain, B., 2001,
  ApJ 546, 20
\bibitem{} Scoccimarro, R., Zaldarriaga, M., Hui, L., 1999, ApJ 527, 1
\bibitem{} Schneider, P., Bartelmann, M., 1995, MNRAS 273, 475
\bibitem{} Seljak, U., 2000, MNRAS 318, 203
\bibitem{} Smith, R. E., et al., 2003, MNRAS 341, 1311
\bibitem{} Springel, V., Yoshida, N., White, S. D. M., 2001, NewA 6,
  79
\bibitem{} Szapudi, I., 2001, in The Onset of Nonlinearity in
  Cosmology, eds. J. N. Fry, J. R. Buchler, H. Kandrup, Annals
of the New York Academy of Sciences 927, p. 94
\bibitem{} Zel'dovich, Ya. B., 1970, A\&A 5, 84
\end{thebibliography}
\end{document}